\documentclass[12pt]{article}
\usepackage{epsfig}
\usepackage{psfrag}
\usepackage{amstext}
\usepackage{latexsym}
\usepackage{indentfirst}
\usepackage{fancyhdr}
\usepackage{amssymb}
\usepackage{textcomp}
\usepackage{latexsym,amssymb,amsmath}
\usepackage{dsfont}
\usepackage{amsfonts}
\usepackage{cite}
\usepackage{bbold}
\usepackage[footnotesize]{caption2}
\usepackage{graphicx}
\usepackage[center,footnotesize,hang]{subfigure}
\usepackage{url}
\usepackage{multirow}
\usepackage{colordvi}
\usepackage{color}
\usepackage{pifont}
\usepackage[mathscr]{eucal}
\usepackage{amsmath}
\usepackage{amsfonts}
\usepackage{footmisc}
\usepackage{ulem}

\catcode`@=11
\def\marginnote#1{}
\newcount\hour
\newcount\minute
\newtoks\amorpm
\hour=\time\divide\hour by60
\minute=\time{\multiply\hour by60 \global\advance\minute by-\hour}
\edef\standardtime{{\ifnum\hour<12 \global\amorpm={am}%
        \else\global\amorpm={pm}\advance\hour by-12 \fi
        \ifnum\hour=0 \hour=12 \fi
        \number\hour:\ifnum\minute<10 0\fi\number\minute\the\amorpm}}
\edef\militarytime{\number\hour:\ifnum\minute<10 0\fi\number\minute}
\def\draftlabel#1{{\@bsphack\if@filesw {\let\thepage\relax
   \xdef\@gtempa{\write\@auxout{\string
      \newlabel{#1}{{\@currentlabel}{\thepage}}}}}\@gtempa
   \if@nobreak \ifvmode\nobreak\fi\fi\fi\@esphack}
        \gdef\@eqnlabel{#1}}
\def\@eqnlabel{}
\def\@vacuum{}
\def\draftmarginnote#1{\marginpar{\raggedright\scriptsize\tt#1}}
\def\draft{\oddsidemargin 0.0truein
        \def\@oddfoot{\sl preliminary draft \hfil
        \rm\thepage\hfil\sl\today\quad\militarytime}
        \let\@evenfoot\@oddfoot \overfullrule 3pt
        \let\label=\draftlabel
        \let\marginnote=\draftmarginnote
   \def\@eqnnum{(\theequation)\rlap{\kern\marginparsep\tt\@eqnlabel}%
\global\let\@eqnlabel\@vacuum}  }
\catcode`@=12
\renewcommand{\thefootnote}{\fnsymbol{footnote}}

\begin{document}
\begin{titlepage}

\begin{flushright}
\end{flushright}

\vskip 1.5cm
\begin{center}
{\Large\bf Dirac Neutrinos with $S_4$ Flavor Symmetry in Warped Extra Dimensions}
\end{center}
\vskip 0.2  cm
\vskip 0.5  cm
\begin{center}
{\large Gui-Jun Ding$^{a}$\footnote{E-mail: {\tt dinggj@ustc.edu.cn}},~~Ye-Ling Zhou$^b$\footnote{E-mail: {\tt zhouyeling@ihep.ac.cn}}}\\[0.4cm]
{\it $^a$ Department of Modern Physics, \\ University of Science and Technology of China, Hefei, Anhui 230026, China}\vspace{0.25cm}\\
\vskip 0.1cm
{\it $^b$ Institute of High Energy Physics, Chinese Academy of Sciences, Beijing 100049, China}
\end{center}
\vskip 0.7cm

\begin{abstract}
\noindent
We present a warped extra dimension model with the custodial symmetry $SU(2)_L\times SU(2)_R\times U(1)_X\times P_{LR}$ based on the flavor symmetry $S_4\times Z_2\times Z'_2$, and the neutrinos are taken to be Dirac particles. At leading order, the democratic lepton mixing is derived exactly, and the high dimensional operators introduce corrections of order $\lambda_c$ to all the three lepton mixing angles such that agreement with the experimental data can be achieved. The neutrino mass spectrum is predicted to be of the inverted hierarchy and the second octant of $\theta_{23}$ is preferred. We suggest the modified democratic mixing, which is obtained by permuting the second and the third rows of the democratic mixing matrix, should be a good first order approximation to understanding sizable $\theta_{13}$ and the first octant of $\theta_{23}$. The constraints on the model from the electroweak precision measurements are discussed. Furthermore, we investigate the lepton mixing patterns for all the possible residual symmetries $G_{\nu}$ and $G_{l}$ in the neutrino and charged lepton sectors, respectively.

\end{abstract}
\end{titlepage}

\def\thefootnote{\arabic{footnote}}
\setcounter{footnote}{0}
\vskip2truecm

\section{\label{sec:introduction}Introduction}
Warped extra dimensions, also known as Randall-Sundrum (RS) models, were first proposed to solve the gauge hierarchy problem in the standard model (SM) \cite{RS_model}. In addition, they provide a novel and powerful framework for understanding flavor physics. The observed SM charged fermion mass hierarchies can be naturally generated due to the overlap of the fermion and Higgs wavefunctions, if the SM fermions and gauge bosons are allowed to propagate in the bulk \cite{bulkfields}.  Indeed, from this approach, both the quark mass hierarchies and the CKM mixing angles can be accurately reproduced as shown in Refs. \cite{hierarchy1_RS,hierarchy2_RS}, and the Yukawa couplings can be naturally of order one with a completely random pattern. However, the electroweak precision data mainly from the Peskin-Takeuchi $T$ parameter turn out to be so restrictive that the masses of the Kaluza-Klein (KK) modes of bulk fields are pushed to somewhat high scales that make their discovery at the LHC extremely challenging \cite{Huber:2000fh}. To reduce the size of the oblique corrections, one can extend the SM gauge symmetry to a bulk custodial symmetry \cite{Agashe:2003zs,Agashe:2006at}, introduce large brane kinetic terms \cite{Davoudiasl:2002ua} or promote the Higgs to a bulk field \cite{Cabrer:2010si}.

In general, the 5D fermion bulk mass parameters, which determine the localization of the zero mode field in the extra dimension, are not degenerate for different fermions in order to account for the mass hierarchy. As a result, dangerously large flavor changing neutral current (FCNC) processes arise already at the tree level through the KK gauge boson exchange \cite{Gherghetta:2000qt,Agashe:2004ay,Cacciapaglia:2007fw, Gedalia:2009ws, Kitano:2000wr,Moreau:2006np,Agashe:2006iy}. To soften the FCNC constraints, one could introduce additional bulk flavor symmetry such as the 5D minimal flavor violation \cite{Fitzpatrick:2007sa,Chen:2008qg} or some dynamical mechanisms \cite{Csaki:2009wc}. Another interesting way is to impose a discrete flavor symmetry to provide extra flavor protection \cite{Csaki:2008qq}, no additional gauge bosons have to be introduced in this setup, and we could possibly obtain realistic lepton masses and flavor mixing pattern through the spontaneous breaking of the discrete flavor symmetry. The idea of combining the discrete flavor symmetry and extra dimension is quite attractive and has already been discussed in the literature within the context of large extra dimensions \cite{Altarelli:2005yp,Altarelli:2008bg}, warped extra dimensions \cite{Chen:2009gy} and holographic composite Higgs models \cite{delAguila:2010vg}. However, all these models try to generate tri-bimaximal neutrino mixing \cite{TBmix} at leading order (LO), where $\theta_{13}=0^{\circ}$, $\theta_{12}\simeq35.3^{\circ}$ and $\theta_{23}=45^{\circ}$. It is possible to generate non-zero $\theta_{13}$ by the next to leading order (NLO) corrections, but in the generic case, the induced deviations should be of the same order for all the three mixing angles. Due to the strong constraint from the accurately measured $\theta_{12}$, these models are expected to give rise to $\theta_{13}$ at most of order $\lambda^2_c$, where $\lambda_c\simeq0.23$ is the Cabibbo angle. Recently the neutrino oscillation experiments T2K \cite{Abe:2011sj}, MINOS \cite{Adamson:2011qu}, Double CHOOZ \cite{Abe:2011fz}, Daya Bay \cite{An:2012eh} and RENO \cite{Ahn:2012nd},
together with the global fitting of mixing parameters \cite{Tortola:2012te,Fogli:2012ua,GonzalezGarcia:2012sz}, confirm a sizeable reactor angle $\theta_{13}\sim9^{\circ}$. Consequently, the tri-bimaximal mixing as a LO approximation is strongly disfavored, unless the underlying theory is capable of providing sufficiently large corrections to $\theta_{13}$ without affecting too much the solar angle.

Flavor symmetry has been widely applied to explaining the structure of the leptonic mixing angles in the past. Roughly speaking, there are two approaches based on flavor symmetry to understand the current lepton mixing parameters, particularly the sizable $\theta_{13}$ and non-maximal $\theta_{23}$. The first one is to construct models that give some new mixing textures, which are in good agreement with the present data and initially admit a non-zero $\theta_{13}$. The so-called Toorop-Feruglio-Hagedorn (TFH) mixing pattern, which gives $\sin^2\theta_{13}=(2-\sqrt{3})/6\simeq0.045$, $\sin^2\theta_{12}=(8-2\sqrt{3})/13\simeq0.349$ and $\sin^2\theta_{23}=(5+2\sqrt{3})/13\simeq0.651$, is a viable substitute for the tri-bimaximal mixing, and it can be naturally produced from the $\Delta(96)$ family symmetry group \cite{Toorop:2011jn,Ding:2012xx,King:2012in}.
Under the hypothesis of Majorana neutrinos~\footnote{They assume that the remnant symmetries in the neutrino and charged lepton sectors are $Z_2\times Z_2$ and $Z_3$, respectively, and the left-handed lepton doublets are assigned to the three dimensional irreducible representation of the flavor symmetry group.}, an extensive general scan of all finite discrete groups with order less than 1536 is performed to obtain the corresponding predictions for lepton mixing angles \cite{Holthausen:2012wt}. Only three groups $\Delta(600)$, $(Z_{18}\times Z_6)\rtimes S_3$ and $\Delta(1536)$ are found to predict lepton mixing angles within $3\sigma$ range of the current global fits. That is to say, the order of the flavor symmetry group is required to be rather large~\footnote{If only $Z_2$ instead of $Z_2\times Z_2$ subgroup is preserved in the neutrino sector, the leptonic mixing angles can not be completely fixed, although they are usually correlated with each other. One can tune the parameters, which are undetermined by the flavor symmetry but are constrained by the observations, to account for the data on neutrino mixing. In this case, the order of the flavor symmetry group can be smaller \cite{Hernandez:2012ra}.} and the resulting flavor models become typically complex \cite{Lam:2013ng}. The second approach is starting from some texture which is outside the current $3\sigma$ range, but the sizable NLO corrections could pull the three mixing angles into the experimentally fovared range. In Ref. \cite{Altarelli:2009gn}, the well-known bimaximal mixing \cite{Barger:1998ta} with $\theta_{13}=0$, $\theta_{12}=\theta_{23}=45^{\circ}$ was taken as the first order approximation, and the model was tactfully constructed such that $\theta_{12}$ and $\theta_{13}$ were corrected by terms of order $\lambda_c$ while $\theta_{23}$ was unchanged at this order. As a result, the mixing angles could be in accordance with the measured values. There are also other attempt of seeking for alternative schemes based upon different flavor symmetries~\cite{Ding:2012wh}. In this work, we shall investigate the democratic (DC) mixing pattern \cite{Fritzsch:1995dj} as the LO mixing pattern. In order to be compatible with experimental data, all the three mixing angles should undergo corrections of order $0.1\sim0.2$, i.e. the NLO corrections should be of order $\lambda_c$, as has been pointed out in Refs. \cite{Xing:2011at,Xing:2012ej}. The corresponding models would be somewhat more natural and more easily constructed than the bimaximal case, since we don't need specific dynamical tricks to obtain the different order of magnitudes of the NLO corrections to the three mixing angles. Note that although many phenomenological studies have been performed for the DC mixing, there still doesn't exist a dynamical model realizing this mixing pattern without fine-tuning.

It is now generally accepted that neutrinos have tiny masses. However, the nature (Dirac vs. Majorana) of neutrinos and thus the origin of neutrino masses remain unknown. Historically in the 4-dimensional (4D) space-time, the preference is that they are of the Majorana type, since the smallness of the neutrino masses can be elegantly explained by the see-saw mechanism \cite{SeeSaw}. The see-saw mechanism is also valid in the 5-dimensional (5D) case \cite{Huber:2003sf,Huber:2002gp}. Moreover, warped extra dimension can generate small Dirac neutrino masses without fine-tuning from the appropriate localization of neutrinos in the bulk \cite{Grossman:1999ra}. In addition, it was argued in \cite{Agashe:2008fe}, that in RS type scenarios, constraints from lepton flavour violation favor Dirac neutrinos over Majorana neutrinos. Therefore we restrict ourselves to the case of Dirac neutrinos. We note that Dirac neutrino in $A_4$ flavor symmetry was recently discussed to generate non-zero $\theta_{13}$~\cite{Memenga:2013vc} from the antisymmetric contraction of the neutrino Dirac coupling at LO. Another motivation for considering Dirac neutrinos is below:  following the bottom-up method presented in Ref. \cite{Lam:2008sh,Ding:2011cm}, we find that there is no discrete group that contains all of the symmetries needed for the DC mixing~\footnote{For convenience, we work in the base in which $\widetilde{m}_l\equiv m_lm^{\dagger}_l$ is diagonal, where $m_l$ is the charged lepton mass matrix. It is easy to see that the symmetry transformation matrix $G_l$, which is determined by the condition $G^{\dagger}_{l}\widetilde{m}_lG_{l}=\widetilde{m}_l$, is a diagonal and non-degenerate $3\times3$ phase matrix. In the case that neutrinos are Majorana particles, the light neutrino mass matrix for DC mixing is of the form $m^{DC}_{\nu}=U^{*}_{DC}\text{diag}(m_1,m_2,m_3)U^{\dagger}_{DC}$. The symmetry transformations $G_{i}$, which satisfy $G^{T}_{i}m^{DC}_{\nu}G_i=m^{DC}_{\nu}$, are determined to be $G_1=+u_1u^{\dagger}_1-u_2u^{\dagger}_2-u_3u^{\dagger}_3$, $G_2=-u_1u^{\dagger}_1+u_2u^{\dagger}_2-u_3u^{\dagger}_3$ and $G_3=-u_1u^{\dagger}_1-u_2u^{\dagger}_2+u_3u^{\dagger}_3$ besides the identity transformation, where $u_i$ is the $i^{th}$ column of $U_{DC}$. They satisfy $G^2_i=1,~~G_iG_j=G_jG_i=G_k(i\neq j\neq k)$. Consequently the symmetry group of the neutrino mass matrix $m^{DC}_{\nu}$ is the Klein four group $K_4\cong Z_2\times Z_2$. Denoting the underlying family symmetry group at high energies as $\mathcal{G}$, then the symmetry transformations $G_l$ and $G_i$ should be the elements of $\mathcal{G}$. In the case of $\mathcal{G}$ being a finite group,
there should be some integers $n$ and $m_i$ such that $G^{n}_{l}=\left(G_{i}G_{l}\right)^{m_i}=1$ with $n\geq3$ which results from the requirement that $G_{l}$ is nondegenerate. We have performed a systematic scan of the possible values of $n$ up to $n=200$, we are unable to find solutions for the integers $m_i$ such that $\left(G_{i}G_{l}\right)^{m_i}=1$, and hence the symmetry groups in these cases are infinite. Therefore we conclude that there is no discrete flavor symmetry group that contains all of the symmetries needed for the DC mixing, although one cannot rule out the possibility of a discrete group with a very large order. This is the reason why the discrete flavor symmetry origin of the DC mixing has not been proposed so far. Note that the $S_{3_L}\times S_{3_R}$ symmetry can immediately lead to the so-called democratic mass matrix in which each matrix element has the same value~\cite{Fritzsch:1989qm}, where $S_{3_L}$ and $S_{3_R}$ are symmetric groups of degree three acting on the left-handed and the right-handed fermion fields respectively. However, the DC mixing can not be uniquely determined by the democratic mass matrix, and in fact only the third row of DC mixing matrix is fixed.} and therefore the DC mixing can not be exactly derived from some discrete flavor symmetry without fine-tuning, if neutrinos are Majorana particles. Within the framework of warped extra dimensions, we shall show that the DC mixing pattern can be naturally produced at LO from $S_4$ flavor symmetry. The model is built in such a way that the DC mixing receives corrections of order $\lambda_c$ from the higher order terms. As a result, all the three leptonic mixing angles are corrected by terms of order $\lambda_c$, and thus, an agreement with the experimental data is achieved. On the other hand, if the LO lepton mixing is chosen to be of the tri-bimaximal form as many previous work, one has to introduce some special dynamics or very tactfully construct the model such that the reactor angle receives much larger subleading corrections than the solar and the atmospheric mixing angles in order to be in accordance with experimental data.

The paper is organized as follows. In section \ref{sec:DC} we investigate the general properties of the Dirac neutrino mass matrix leading to the DC mixing in the flavor basis. In section \ref{sec:model} we construct our model in the context of custodial warped extra dimensions based on $S_4$ flavor symmetry, where the Higgs field is assumed to propagate in the bulk and the DC mixing pattern is realized in the LO. In section \ref{sec:NLO} the corrections to the DC mixing angles from higher dimensional operators are analyzed. In section \ref{sec:EWC} the constraints from electroweak precision measurements are studied. In section \ref{sec:residual} we explore all the possible leptonic mixing patterns within $S_4$ by scanning the possible $G_{\nu}$ and $G_{l}$ subgroups preserved in the neutrino and charged lepton sectors, respectively. Finally section \ref{sec:conclusion} is devoted to our conclusion and summary.

\section{\label{sec:DC}Democratic mixing}

In a specific phase convention, the democratic (DC) mixing  matrix is given by \cite{Fritzsch:1995dj}
\begin{equation}
\label{eq:UDC}U_{DC}=\left(\begin{array}{ccc}
\frac{i}{\sqrt{2}}   &   -\frac{i}{\sqrt{2}}   &  0 \\
-\frac{1}{\sqrt{6}}   &  -\frac{1}{\sqrt{6}}   &  \sqrt{\frac{2}{3}}   \\
\frac{1}{\sqrt{3}}    &   \frac{1}{\sqrt{3}}   & \frac{1}{\sqrt{3}}
\end{array}
\right)\,.
\end{equation}
It gives rise to the three leptonic mixing angles as follows
\begin{equation}
\theta^{DC}_{12}=45^{\circ},\qquad\quad \theta^{DC}_{23}=\arctan\sqrt{2}\simeq54.7^{\circ},\qquad\quad  \theta^{DC}_{13}=0^{\circ}  \,.
\end{equation}
The DC mixing is closely related to the well-known tri-bimaximal mixing \cite{TBmix}, the former can be obtained by transposing the latter one and then permuting its rows and columns respectively:
\begin{equation}
U_{DC}=\left(\begin{array}{ccc}
1  & 0  &  0\\
0  & 0  &  1  \\
0  & 1  &  0
\end{array}\right) \left(\begin{array}{ccc}
0  & 0  &  1 \\
0  & 1  &  0 \\
1  & 0  &  0
\end{array}\right) U^{T}_{TB}\left(\begin{array}{ccc}
0  & 0  &  1\\
0  & 1  &  0 \\
1  & 0  &  0
\end{array}\right)\,,
\end{equation}
where $U_{TB}$ is the tri-bimaximal mixing matrix
\begin{equation}
U_{TB}=\left(\begin{array}{ccc}
\sqrt{\frac{2}{3}}    & \frac{1}{\sqrt{3}}  &  0  \\
-\frac{1}{\sqrt{6}}   &  \frac{1}{\sqrt{3}} & -\frac{i}{\sqrt{2}}  \\
-\frac{1}{\sqrt{6}}   &  \frac{1}{\sqrt{3}} & \frac{i}{\sqrt{2}}
\end{array}\right)\,.
\end{equation}
In the basis where the charged lepton mass matrix is diagonal, the hermitian combination of the neutrino mass matrix $\mathcal{M}_{\nu}\equiv m_{\nu}m^{\dagger}_{\nu}$ for the DC mixing is given by
\begin{eqnarray}
\nonumber\hskip-0.2in\mathcal{M}_{\nu}&=&U_{DC}\;\mathrm{ diag}(m^2_1,m^2_2,m^2_3)U^{\dagger}_{DC}  \\
\label{eq:5}&=&\frac{m^2_1}{6}
\begin{pmatrix}
 3 & -\sqrt{3}\,i & \sqrt{6}\,i \\
 \sqrt{3}\,i & 1 & -\sqrt{2} \\
 -\sqrt{6}\,i & -\sqrt{2} & 2
\end{pmatrix}+\frac{m^2_2}{6}\begin{pmatrix}
 3 &  \sqrt{3}\,i & -\sqrt{6}\,i \\
 -\sqrt{3}\,i & 1 & -\sqrt{2} \\
 \sqrt{6}\,i & -\sqrt{2} & 2
\end{pmatrix}+\frac{m^2_3}{3}\begin{pmatrix}
0 & 0 & 0 \\
 0 & 2 & \sqrt{2} \\
 0 & \sqrt{2} & 1
\end{pmatrix}\,,
\end{eqnarray}
where $m_{1,2,3}$ are the light neutrino masses. As a result, the most general mass matrix $\mathcal{M}_{\nu}$ corresponding to DC mixing is of the form
\begin{equation}
\label{eq:6}\mathcal{M}_{\nu}=\left(
\begin{array}{ccc}
 x & i y & -i\sqrt{2}\; y \\
 -i y & z & (z-x)/\sqrt{2} \\
 i \sqrt{2}\; y & (z-x)/\sqrt{2} &
   \left(x+z\right)/2
\end{array}
\right)\,,
\end{equation}
where the parameters $x$, $y$ and $z$ are real. The squared light neutrino masses can be expressed in terms of $x$, $y$ and $z$ as
$m^2_1=x-\sqrt{3}\;y$, $m^2_2=x+\sqrt{3}\;y$ and $m^2_3=(3z-x)/2$. If we perform a  unitary transformation $\nu_{L}\rightarrow G \nu_L$, where $\nu_L=(\nu_e,\nu_{\mu},\nu_{\tau})_L^T$ is the left-handed neutrino field, then $\mathcal{M}_{\nu}$ transforms as $\mathcal{M}_{\nu}\rightarrow G^{\dagger}\mathcal{M}_{\nu}G$. The hermitian combination $\mathcal{M}_{\nu}$ in Eq. (\ref{eq:5}) or Eq. (\ref{eq:6}) for the DC mixing would be invariant, if the transformation $G$ takes the form
\begin{eqnarray}
\label{eq:dc_symm1}&&G_1(\alpha,\beta)=\frac{1}{6}\left(
\begin{array}{ccc}
 3 \left(e^{i \alpha}+e^{i \beta}\right) & -i \sqrt{3}
   \left(e^{i \alpha}-e^{i \beta}\right) & i \sqrt{6}
   \left(e^{i \alpha}-e^{i \beta}\right) \\
 i \sqrt{3} \left(e^{i \alpha}-e^{i \beta}\right) &
   e^{i \alpha}+5 e^{i \beta} & \sqrt{2} \left(e^{i
   \beta}-e^{i \alpha}\right) \\
 -i \sqrt{6} \left(e^{i \alpha}-e^{i \beta}\right) &
   \sqrt{2} \left(e^{i \beta}-e^{i \alpha}\right) & 2
   \left(e^{i \alpha}+2 e^{i \beta}\right)
\end{array}
\right)\,,
\end{eqnarray}
or
\begin{eqnarray}
\label{eq:dc_symm2}&&G_{2}(\alpha,\beta)=\frac{1}{6}\left(
\begin{array}{ccc}
 3 \left(e^{i \alpha}+e^{i \beta}\right) & i \sqrt{3}
   \left(e^{i \alpha}-e^{i \beta}\right) & -i \sqrt{6}
   \left(e^{i \alpha}-e^{i \beta}\right) \\
 -i \sqrt{3} \left(e^{i \alpha}-e^{i \beta}\right) &
   e^{i \alpha}+5 e^{i \beta} & \sqrt{2} \left(e^{i
   \beta}-e^{i \alpha}\right) \\
 i \sqrt{6} \left(e^{i \alpha}-e^{i \beta}\right) &
   \sqrt{2} \left(e^{i \beta}-e^{i \alpha}\right) & 2
   \left(e^{i \alpha}+2 e^{i \beta}\right)
\end{array}
\right)\,,
\end{eqnarray}
where $\alpha$ and $\beta$ are arbitrary real parameters. It is straightforward to check that the above symmetry transformations satisfy the following multiplication rules
\begin{eqnarray}
\nonumber&&G_{1}(\alpha_1,\beta_1)G_{1}(\alpha_2,\beta_2)=G_{1}(\alpha_1+\alpha_2,\beta_1+\beta_2) \,, \\
\nonumber&&G_{2}(\alpha_1,\beta_1)G_{2}(\alpha_2,\beta_2)=G_{2}(\alpha_1+\alpha_2,\beta_1+\beta_2) \,, \\
&&G_{1}(\alpha_1,\beta_1)G_{2}(\alpha_2,\beta_2)=G_{2}(\alpha_2,\beta_2)G_{1}(\alpha_1,\beta_1)\,.
\end{eqnarray}
Therefore, the symmetry group of $\mathcal{M}_{\nu}$, which is generated by the symmetry transformations $G_{1}(\alpha,\beta)$ and $G_2(\alpha,\beta)$, is isomorphic to $U(1)\times U(1)\times U(1)\times U(1)\equiv \mathcal{G}_{\nu}$. Generally the symmetry group of the Dirac neutrino mass matrix for mass independent mixing patterns is $\mathcal{G}_{\nu}\equiv U(1)\times U(1)\times U(1)\times U(1)$ in the charged lepton diagonal base, although the specific forms of the symmetry transformations $G_{1}(\alpha,\beta)$ and $G_{2}(\alpha,\beta)$ depend on the mixing texture. On the other hand, if the hermitian combination of the neutrino mass matrix $\mathcal{M}_{\nu}=m_{\nu}m^{\dagger}_{\nu}$ is invariant under the $\mathcal{G}_{\nu}$ symmetry group generated by $G_{1}(\alpha,\beta)$ and $G_{2}(\alpha,\beta)$ in Eqs. (\ref{eq:dc_symm1}) and (\ref{eq:dc_symm2}), it would be diagonalized by the DC mixing matrix precisely. In fact, it is sufficient to require $\mathcal{M}_{\nu}$ invariant under one element $h\in\mathcal{G}_{\nu}$ with non-degenerate eigenvalues for $\mathcal{M}_{\nu}$ being diagonalized by DC mixing. For example, $\mathcal{M}_{\nu}\equiv m_{\nu}m^{\dagger}_{\nu}$ invariant under an order three element $h$ with
\begin{eqnarray}
h=G_1(4\pi/3,0)G_2(2\pi/3,0)=\frac{1}{2}\left(
\begin{array}{ccc}
-1&-1& \sqrt{2}  \\
1 & 1& \sqrt{2} \\
-\sqrt{2}  & \sqrt{2}  &0
\end{array}
\right)\,,
\end{eqnarray}
would be diagonalized by the DC mixing matrix. We note that this unitary transformation of $h$ is exactly the representation matrix for the $S_4$ generator $T$ in three dimensional irreducible representations, as presented in the Appendix A. This is the reason why we can reproduce DC mixing exactly at LO from the finite $S_4$ family symmetry group, as will be shown in the following. It is not necessary that the flavor symmetry group is large enough to include the left-handed (LH) neutrino symmetry group $\mathcal{G}_{\nu}\equiv U(1)\times U(1)\times U(1)\times U(1)$ as a subgroup so that it can be preserved in the neutrino sector. On the contrary, if neutrinos are Majorana particles, generally the neutrino mass matrix symmetry group $K_4\cong Z_2\times Z_2$ should be a subgroup of the full flavor symmetry group, or one $Z_2$ belongs to the flavor symmetry group while the other is accidental.

\section{\label{sec:model}The structure of the model}

In this section, we present a warped extra dimension model of leptons based on $S_4\times Z_2\times Z'_2$ flavor symmetry. The LO lepton mixing is of the DC form in this model, and its phenomenological implications will be discussed.

\subsection{The model}
\begin{table}[t!]
\begin{center}
\begin{tabular}{|c|c|c|c||c|c|c|}\hline\hline
 ${\tt Fields}$     &    $~SU(2)_L~$     &   $~SU(2)_R~$       &      $~U(1)_X~$     &   $~S_4~$       &     $~Z_2~$        &     $~Z'_2~$   \\ \hline

$\xi^{i}_1$   &   $\mathbf{2}$      &     $\mathbf{2}$      &    0       &    $\mathbf{3}$     &   $+$       &   $+$   \\

$\xi^{i}_2$   &    $\mathbf{1}$     &     $\mathbf{1}$      &    0       &    $\mathbf{3}$     &   $+$       &   $-$   \\ \hline

$T^{1}_3(T^{1}_4)$  &  \multirow{3}{*}{$\mathbf{3}(\mathbf{1})$}    &    \multirow{3}{*}{$\mathbf{1}(\mathbf{3})$}    &   \multirow{3}{*}{0}   &   $\mathbf{1}$   &   $-$   &   $+$   \\
$T^{2}_3(T^{2}_4)$  &                                               &                                                 &                        &   $\mathbf{1'}$  &   $-$   &   $+$   \\
$T^{3}_3(T^{3}_4)$  &                                               &                                                 &                        &   $\mathbf{1}$   &   $+$   &   $+$   \\ \hline

$H$                 &          $\mathbf{2}$                         &            $\mathbf{2}$                         &      0                 &   $\mathbf{1}$   &    $+$  &   $+$  \\ \hline\hline

$\chi(IR)$          &       \multirow{5}{*}{$\mathbf{1}$}           &    \multirow{5}{*}{$\mathbf{1}$}          &     \multirow{5}{*}{0}         &    $\mathbf{3}$  &   $+$  &  $+$   \\
$\varphi(IR)$       &                                               &                                           &                                &    $\mathbf{3}'$ &   $-$  &  $+$   \\

$\zeta(UV)$         &                                               &                                           &                                &    $\mathbf{1}$  &   $+$   &  $-$  \\
$\phi(UV)$          &                                               &                                           &                                &    $\mathbf{3}'$ &   $+$  &   $-$  \\
$\rho(UV)$          &                                               &                                           &                                &    $\mathbf{2}$  &   $+$  &   $+$  \\ \hline\hline

\end{tabular}
\caption{\label{tab:field} Representation assignments for the bulk fields and the localized scalars (i.e. flavon fields) under the gauge symmetry $SU(2)_L\times SU(2)_R\times U(1)_X\times P_{LR}$ and the flavor symmetry group $S_4\times Z_2\times Z'_2$, where the superscripts are the family indices. }
\end{center}
\end{table}

Our model is formulated on a slice of $\text{AdS}_5$ space. The background metric is of the usual Randall-Sundrum form \cite{RS_model}
\begin{equation}
ds^2=e^{-2ky}\eta_{\mu\nu}dx^{\mu}dx^{\nu}-dy^2\,,
\end{equation}
where $\eta_{\mu\nu}=\text{diag}(1,-1,-1,-1)$, the fifth coordinate is restricted to the interval $0\leq y\leq L$, and $k\sim\mathcal{O}(M_{Pl})$ is the $\text{AdS}_5$ curvature scale. As usual the UV (Planck) brane is localized at $y=0$, and the IR (TeV) brane is at $y=L$. It is useful to define a parameter $k'=ke^{-kL}$ for the following discussions. To solve the gauge hierarchy problem, $k'$ has to be at the TeV scale. The SM electroweak gauge group is extended to
\begin{equation}
G_{\text{bulk}}=SU(2)_L\times SU(2)_R\times U(1)_X\times P_{LR}\,,
\end{equation}
in the bulk. The 5D gauge coupling constants of $SU(2)_L$, $SU(2)_R$ and $U(1)_X$ are $g_{5L}$, $g_{5R}$ and $g_{5X}$, respectively. The presence of $SU(2)_R$ implies an unbroken custodial symmetry which eliminates the excessively large contributions to the $T$ parameter due to the Kaluza-Klein (KK) excitations of the bulk fields. The discrete symmetry $P_{LR}$, interchanging the two $SU(2)$ groups, is proposed to protect the SM $Z$ boson couplings to the LH down-type quarks, in particular the $Zb_{L}\bar{b}_L$ coupling, from large corrections \cite{Blanke:2008zb}. Clearly the $P_{LR}$ symmetry implies that the gauge couplings of $SU(2)_L$ and $SU(2)_R$ have to be equal $g_{5L}=g_{5R}\equiv g_5$. The bulk gauge symmetry $G_{\text{bulk}}$ is broken to the SM gauge group $SU(2)_L\times U(1)_Y$ on the UV brane by means of the boundary conditions of the electroweak (EW) gauge bosons. The resulting 5D $U(1)_Y$ gauge coupling constant is given by $g^\prime_{5}=g^{}_{5X}/\sqrt{g^2_{5}+g^2_{5X}}$. In order to realize the standard EW symmetry breaking, we introduce a bulk Higgs field which is a self-dual bidoublet of $SU(2)_L\times SU(2)_R$ and can be parameterized as:
\begin{equation}
\label{eq:higgs}H=\frac{1}{2}\left(\begin{array}{cc}
\sqrt{2}\,\pi^{+}   &  -h^0+i\pi^0  \\
h^0+i\pi^0    & \sqrt{2}\,\pi^{-}
\end{array}\right)\,.
\end{equation}
It is a singlet under $U(1)_X$ with $Q_{X}(H)=0$. In the notation of Eq. (\ref{eq:higgs}), $SU(2)_L$ acts vertically while $SU(2)_R$ acts horizontally,
\begin{equation}
H\rightarrow U_LHU^{T}_R,\quad U_L\in SU(2)_L,\quad U_R\in SU(2)_R\,.
\end{equation}
The Goldstone bosons eaten by the gauge bosons are denoted by $\pi^{\pm}$ and $\pi^{0}$. $h^0$ is the physical Higgs boson, and it develops a 4D effective vacuum expectation value (VEV) near the IR brane, which eventually leads to EW symmetry breaking. In order to yield a light Higgs zero mode, the whole bidoublet $H$ has to obey the (++) boundary conditions (BCs), where ``+'' stands for a Neumann boundary condition. Performing the KK decomposition of the Higgs field $H(x^{\mu},y)$, we have
\begin{equation}
H(x^{\mu},y)=\frac{1}{\sqrt{L}}H(x^{\mu})h(y)+\text{heavy KK modes}\,,
\end{equation}
and we can choose the BCs such that its profile $h(y)$ takes the simple form \cite{Cacciapaglia:2006mz}~\footnote{The bulk equations of motion admit two solutions for the profile of the bulk Higgs: one is proportional to $e^{(2-\beta)k(y-L)}$ and the other is proportional to $e^{(2+\beta)k(y-L)}$. We could choose the BCs to select one of the above two solutions. In the present work, we take the first one in order to generate light neutrino masses of correct order of magnitude, as will be discussed in the following. }
\begin{equation}
\label{eq:higgs_profile}h(y)=\sqrt{\frac{2(1-\beta)kL} {1-e^{-2(1-\beta)kL}}}\,e^{kL}e^{(2-\beta)k(y-L)}\,.
\end{equation}
The parameter $\beta$ is related to the Higgs bulk mass $m_H$ by $\beta=\sqrt{4+m^2_H/k^2}$, which controls the localization of the profile of the VEV in the bulk. In order to localize the Higgs VEV close to the IR brane such that the electroweak symmetry is broken at the TeV scale, the bulk mass squared $m^2_H$ should be negative but it is bounded by $m^2_H\geq-4k^2$ from both the Breitenlohner-Freedman bound \cite{Breitenlohner:1982jf} and the requirement that the dual theory resolves the gauge hierarchy problem \cite{Luty:2004ye}. Furthermore, the VEV of the Higgs zero mode is of the form
\begin{equation}
\langle H(x^{\mu})\rangle=\left(\begin{array}{cc}
0  & -v/2 \\
v/2 &0
\end{array}
\right)\,,
\end{equation}
where $v$ denotes the effective 4D VEV of the zero mode of $h^0$. As has been shown above, a novel feature of the bulk Higgs is that the Higgs VEV has a profile which extends into the bulk and is no longer a constant. This provides new model-building possibilities, and it is relevant to neutrino mass generation in the present work. Moreover, a number of existing studies on bulk Higgs scenarios \cite{Cacciapaglia:2006mz,Davoudiasl:2005uu} demonstrate that the bulk Higgs can improve the naturalness of the warped models and lead to better agreement with the experiment data.

The matter fields of the model transform in the non-minimal representations under $SU(2)_{L}\times SU(2)_R$ \cite{Agashe:2006at,Carena:2006bn}. There are three multiplets per generation ($i=1,2,3$):
\begin{eqnarray}
\nonumber&&\hskip-0.3in\left(\xi^{i}_1\right)_{a\alpha}=\left(\begin{array}{cc}
\chi^{\nu_i}(-+)_{1}   &   \ell^{\nu_i}(++)_0 \\
\chi^{e_i}(-+)_0    &   \ell^{e_i}(++)_{-1}
\end{array}\right)\sim(\mathbf{2},\mathbf{2})_0,\qquad  \xi^{i}_2=\nu^{i}(--)_{0}\sim(\mathbf{1},\mathbf{1})_0,\\
\label{eq:matter_fields}&&\hskip-0.3in\xi^{i}_3=\left(T^i_3\right)_a\oplus\left(T^{i}_4\right)_{\alpha}=\left(\begin{array}{c}
\lambda^{\prime i}(+-)_{1}  \\
N^{\prime i}(+-)_{0}  \\
L^{\prime i}(+-)_{-1}
\end{array}
\right)\oplus\left(\begin{array}{c}
\lambda^{\prime\prime i}(+-)_{1}  \\
N^{\prime\prime i}(+-)_{0} \\
L^{i}(--)_{-1}
\end{array}\right)\sim(\mathbf{3},\mathbf{1})_{0} \oplus(\mathbf{1},\mathbf{3})_{0}\,,
\end{eqnarray}
where $SU(2)_L$ indices are denoted by Latin letters, and $SU(2)_R$ indices are denoted by Greek letters. The subscripts of the various components indicate their electric charges, and the $U(1)_X$ charges of all the above leptonic matter fields are vanishing. Note that in the notation of Eq. (\ref{eq:matter_fields}), the matter fields transform in the following way under the gauge transformations $U_{L}\in SU(2)_L$ and $U_{R}\in SU(2)_R$,
\begin{equation}
\xi^{i}_1\rightarrow U_L\xi^{i}_1U^{T}_R\,,\qquad T^{i}_3\rightarrow U_LT^{i}_3U^{\dagger}_L\,,\qquad T^{i}_4\rightarrow U_{R}T^{i}_4U^{\dagger}_R\,.
\end{equation}
Here we have denoted  $T^{i}_3=\tau^{a}\left(T^{i}_3\right)_a$ and $T^{i}_4=\tau^{\alpha}\left(T^{i}_4\right)_{\alpha}$, where $\tau^{a}=\sigma^{a}/2$ and $\tau^{\alpha}=\sigma^{\alpha}/2$ are the generators of the fundamental $SU(2)_L$ and $SU(2)_R$ representations  respectively with $\sigma^{a}$ and $\sigma^{\alpha}$ being the Pauli matrices. The signs in brackets in Eq. (\ref{eq:matter_fields}) indicate the
BCs for the LH fermion mode on the UV and IR branes, respectively, where $``+"$ denotes a modified Neumann BC and $``-"$ stands for a Dirichlet BC. The corresponding right-handed (RH) modes, which are necessarily present in a 5D theory, obey opposite BCs. As only fields with $(++)$ BCs have massless zero modes, up to small mixing effects with other massive modes due to the transformation to mass eigenstates, the low energy spectrum will contain a LH doublet $(\ell^{\nu_i(0)}_L , \ell^{e_i(0)}_L)$ and two RH singlets $\nu^{i(0)}_R$ and $L^{i(0)}_R$ $(i=1,2,3)$ for each lepton generation. The field contents of the SM are reproduced exactly.

The full flavor symmetry of the model is $S_4\times Z_2\times Z'_2$. The three generations $\xi^{i}_1\;(i=1,2,3)$, which contain LH zero modes $(\ell^{\nu_i(0)}_L , \ell^{e_i(0)}_L)$ transforming as an $SU(2)_L$ doublets, are embedded into an $S_4$ triplet $\mathbf{3}$. The three generations of $\xi^{i}_2\;(i=1,2,3)$ containing RH neutral singlets $\nu^{i(0)}_R$, are also assigned to form an $S_4$ triplet $\mathbf{3}$. While each generation of $\xi^{i}_3\;(i=1,2,3)$, which allows a RH singlet $L^{i(0)}_R$ with electric charge $-1$, transforms as nonequivalent one-dimensional representation $\mathbf{1}$ or $\mathbf{1}'$~\footnote{$S_4$ group has only two nonequivalent  one-dimensional representations $\mathbf{1}$ and $\mathbf{1}'$, consequently $S_4$ can not differentiate the three fields $\xi^{1,2,3}_3$ completely.}. The auxiliary symmetry $Z_2$ is imposed to further distinguish the three generation $\xi^{1,2,3}_3$ fields, and the purpose of the other $Z'_2$ is to eliminate certain dangerous operators that would otherwise contribute to the neutrino masses at LO.

The flavor symmetry is broken by a set of flavon fields, which are all singlets under the bulk gauge group and localized on the two branes. Concretely, the flavons $\chi$ and $\varphi$, transforming as $S_4$ triplets $\mathbf{3}$ and $\mathbf{3}'$ on the IR brane, respectively, are responsible for the flavor symmetry breaking in the charged lepton sector. The scalars $\zeta\sim\mathbf{1}$, $\phi\sim\mathbf{3}'$ and $\rho\sim\mathbf{2}$ are localized on the UV brane, and they break the $S_4$ flavor symmetry in the neutrino sector. The field contents of the model and their transformation properties under the gauge and flavor symmetry groups are summarized in Table \ref{tab:field}. By putting the $S_4$ breaking fields $\chi$ and $\varphi$ on the IR brane, and the remaining flavons $\zeta$, $\phi$ and $\rho$ on the UV brane, the questions regarding vacuum alignment are eliminated, as the two sets of flavons are completely sequestered \cite{Altarelli:2005yp}. As usually done in such models, we shall assume that the flavor symmetry $S_4\times Z_2\times Z'_2$ is spontaneously broken by the VEVs of the flavons on the branes as follows
\begin{eqnarray}
\nonumber& \langle\chi\rangle=(0,0,1)v_{\chi}\,,\qquad \langle\varphi\rangle=(0,1,0)v_{\varphi}\,, \\
\label{eq:vacuum}&\langle\zeta\rangle=v_{\zeta}\,,\qquad \langle\phi\rangle=(0,\sqrt{2},1)v_{\phi}\,,\qquad \langle\rho\rangle=(1,0)v_{\rho}\,.
\end{eqnarray}
As shown in Appendix B, this vacuum configuration can be produced from the minimization of the most general renormalizable flavon potential invariant under the imposed symmetry. In the chosen basis which is presented in the Appendix A, it is straightforward to check that we have $TST\langle\chi\rangle=S^2\langle\chi\rangle= TSTS^2\langle\chi\rangle=\langle\chi\rangle$ and $TST\langle\varphi\rangle=-S^2\langle\varphi\rangle= -TSTS^2\langle\varphi\rangle=\langle\varphi\rangle$. Therefore the VEV of $\chi$ breaks the flavor symmetry $S_4$ to its subgroup $K_4=\left\{1,\; TST,\;S^2,\;TSTS^2 \right\}$, while the VEV of $\varphi$ breaks $S_4$ to $Z_2$. On the other hand, these are the most general vacuum configurations which preserve the above $K_4$ and $Z_2$ subgroups, respectively. In the same manner, we have $T\langle\phi\rangle=\langle\phi\rangle$ and likewise it is the most general VEVs invariant under the action of $T$. Consequently, the vacuum alignment of $\zeta$ and $\phi$ breaks $S_4$ to its $Z_3$ subgroup generated by $T$.

In the warped extra dimension, the fermion mass hierarchies are naturally explained through wavefunction localization, which is determined by the bulk fermion mass. The bulk mass Lagrangian for the matter fields $\xi^{i}_1$, $\xi^{i}_2$ and $\xi^{i}_3$, invariant under the flavor symmetry, reads
\begin{eqnarray}
\nonumber\mathcal{L}_m&=&-\sqrt{G}\;\bigg\{\sum^{3}_{i=1}c_{l}k(\overline{\xi}^{i}_1)_{a\alpha}(\xi^{i}_1)_{a\alpha}+\sum^{3}_{i=1}c_{\nu}k\overline{\xi}^{i}_2\xi^{i}_2
+c_{e}k\left[(\overline{T}^{1}_3)_a(T^1_3)_a+(\overline{T}^{1}_4)_{\alpha}(T^{1}_4)_{\alpha}\right]\\
&&+c_{\mu}k\left[(\overline{T}^{2}_3)_a(T^2_3)_a+(\overline{T}^{2}_4)_{\alpha}(T^{2}_4)_{\alpha}\right]
+c_{\tau}k\left[(\overline{T}^{3}_3)_a(T^3_3)_a+(\overline{T}^{3}_4)_{\alpha}(T^{3}_4)_{\alpha}\right]\bigg\}\,,
\end{eqnarray}
where $G=e^{-8ky}$ is the determinant of the RS-metric, the summation over repeated indices is understood, and again $SU(2)_L$ indices are denoted by Latin letters while $SU(2)_R$ indices are denoted by Greek letters. The bulk Dirac masses are conveniently parameterized in terms of the AdS curvature $k$. In general, the bulk mass would be a hermitian $3\times3$ matrix. Once we impose the $S_4$ flavor symmetry in the bulk of the theory, the three generations of the fermion bidoublets $\xi^i_1$ will share one common $c-$parameter which we label by $c_{l}$, since they are embedded into a triplet $\mathbf{3}$ of $S_4$. In the same way, the $S_4$ flavor symmetry implies a family independent bulk mass parameter $c_{\nu}$ for the three gauge singlets $\xi^{i}_2$, which are also assigned to a triplet $\mathbf{3}$. The three generations of $\xi^i_3=T^{i}_3\oplus T^{i}_4$ transform as $\mathbf{1}$, $\mathbf{1}'$ and $\mathbf{1}$, respectively, and thus there are three separate $c-$parameters $c_e$, $c_{\mu}$ and $c_{\tau}$ for these fields. Note that the $P_{LR}$ symmetry imposes a common bulk mass parameter for $T^{i}_3$ and $T^{i}_4$ within the same generation.

The most general Yukawa interactions compatible with both gauge and  flavor symmetries, can be written as
\begin{eqnarray}
\nonumber\mathcal{L}_{IR}&=&-2\sqrt{G}\;\delta(y-L)\bigg\{\frac{y_{\tau}}{\Lambda^{3/2}}\big[\text{Tr}\left(\overline{\xi}_1T^{3}_3H\right)+\text{Tr}\left(\overline{\xi}_1HT^{3T}_4\right)\big]\frac{\chi}{\Lambda}
+\frac{y_{\mu}}{\Lambda^{3/2}}\big[\text{Tr}\left(\overline{\xi}_1T^{2}_3H\right)\\
\label{eq:Yukawa_charged}&&+\text{Tr}\left(\overline{\xi}_1HT^{2T}_4\right)\big]\frac{\varphi}{\Lambda}+\frac{y_{e}}{\Lambda^{3/2}}\big[\text{Tr}\left(\overline{\xi}_1T^{1}_3H\right)+\text{Tr}\left(\overline{\xi}_1HT^{1T}_4\right)\big]\frac{\chi\varphi}{\Lambda^{2}}
\bigg\}+\text{h.c.}+\ldots\,, \\
\label{eq:Yukawa_neutrino}\mathcal{L}_{UV}&=&\sqrt{2G}\;\delta(y)\left[\frac{y_1}{\Lambda^{3/2}}\text{Tr}\left(\overline{\xi}_1H\xi_2\right)\frac{\zeta}{\Lambda}+\frac{y_2}{\Lambda^{3/2}}\text{Tr}\left(\overline{\xi}_1H\xi_2\right)\frac{\phi}{\Lambda}\right]+\text{h.c.}+\ldots\,,
\end{eqnarray}
where the dots stand for higher dimensional operators that will be addressed later in the paper, and all the above operators should be contracted into $S_4$ singlet with the help of Clebsch-Gordon coefficients listed in the Appendix A. All the SM fields would be identified with the 4D components of the zero modes in the KK decomposition of the bulk fields in the following, and this is the so-called zero mode approximation (ZMA).
In general, the mixings between the zero modes of the leptons and their KK excitations could be induced by the Yukawa couplings in Eqs. (\ref{eq:Yukawa_charged}) and (\ref{eq:Yukawa_neutrino}). However, such a mixing effect is not significant, and it is at most of order $m_{\tau}/m_{KK}\sim 10^{-3}$ in rough estimate. Many phenomenological analyses have demonstrated that the effects of the fermionic KK modes are really subleading and therefore can be neglected \cite{Blanke:2008zb,Buras:2009ka}. Once the electroweak and $S_4$ flavor symmetries are spontaneously broken, both neutrinos and charged leptons acquire masses via the above Yukawa interactions. The size of the lepton masses is determined by the wavefunction overlap of lepton zero modes with opposite chirality and the VEV profile of the Higgs field. The IR boundary terms in Eq. (\ref{eq:Yukawa_charged}), which connect $\xi^{i}_1$ with $\xi^{j}_3$, are responsible for the charged lepton masses. Inserting the flavon VEVs shown in Eq. (\ref{eq:vacuum}), we find that the charged lepton mass matrix is exactly diagonal with
\begin{eqnarray}
\nonumber&&m_e=-\frac{y_e}{\left(\Lambda L\right)^{3/2}}\frac{v_{\chi}v_{\varphi}}{\Lambda^{2}}\frac{v}{\sqrt{2}}F(y,\beta,c_{l},c_e)\bigg|_{y=L} \,, \\
\nonumber&&m_{\mu}=\frac{y_{\mu}}{\left(\Lambda L\right)^{3/2}}\frac{v_{\varphi}}{\Lambda}\frac{v}{\sqrt{2}}F(y,\beta,c_{l},c_{\mu})\bigg|_{y=L}\,,\\
&&m_{\tau}=\frac{y_{\tau}}{\left(\Lambda L\right)^{3/2}}\frac{v_{\chi}}{\Lambda}\frac{v}{\sqrt{2}}F(y,\beta,c_{l},c_{\tau})\bigg|_{y=L} \,,
\end{eqnarray}
where we introduce the overlap function $F(y,\beta,c_L,c_R)$ for convenience
\begin{eqnarray}
\nonumber&&\hskip-0.54in F(y,\beta,c_L,c_R)=e^{-ky}h(y)f^{(0)}_L(y,c_L)f^{(0)}_R(y,c_R)\\
&&\hskip-0.43in=\sqrt{\frac{2\left(1-\beta\right)\left(1-2c_L\right)\left(1+2c_R\right)}{\left[1-e^{-2(1-\beta)kL}\right]\left[e^{(1-2c_L)kL}-1\right]\left[e^{(1+2c_R)kL}-1\right]}}\;\;\left(kL\right)^{3/2}e^{-(1-\beta)kL}e^{(2-\beta-c_L+c_R)ky}
\,.
\end{eqnarray}
Here $f^{(0)}_L(y,c_L)$ is the profile of the LH fermionic zero mode, with the form \cite{Grossman:1999ra,Gherghetta:2000qt}
\begin{equation}
f^{(0)}_L(y,c_L)=\sqrt{\frac{(1-2c_L)kL}{e^{(1-2c_L)kL}-1}}\; e^{(\frac{1}{2}-c_L)ky}\,.
\end{equation}
It exists only for (++) BCs for the LH mode. $f^{(0)}_R(y,c_R)$ is the profile of the RH zero mode, which is given by
\begin{equation}
f^{(0)}_R(y,c_R)=\sqrt{\frac{(1+2c_R)kL}{e^{(1+2c_R)kL}-1}}\; e^{(\frac{1}{2}+c_R)ky}\,.
\end{equation}
Here we have performed KK decompositions for fermions as
\begin{equation}
\psi_{L,R}(x^\mu,y,c_{L,R})=\frac{e^{\frac{3}{2}ky}}{\sqrt{L}} \sum^\infty_{n=0}\psi^{(n)}_{L,R}(x^\mu)f^{(n)}_{L,R}(y,c_{L,R})\,.
\end{equation}
Then straightforwardly, we have
\begin{eqnarray*}
F(y=L,\beta,c_L,c_R)=\left(kL\right)^{3/2}\sqrt{\frac{2\left(1-\beta\right) \left(1-2c_L\right)\left(1+2c_R\right)}{\left[1-e^{-2(1-\beta)kL}\right] \left[e^{(1-2c_L)kL}-1\right]\left[e^{(1+2c_R)kL}-1\right]}}\;\; e^{\left(1-c_L+c_R\right)kL}\,,
\end{eqnarray*}
which weakly depends on the Higgs localization parameter $\beta$. The charged lepton mass hierarchies are mainly governed by the exponential factor $e^{\left(1-c_L+c_R\right)kL}$ in the usual way. Note that the flavor symmetry breaking provides an additional suppression factor $v_{\varphi}/\Lambda$ ($v_{\chi}/\Lambda$) for the electron mass with respect to the tau (muon) case.

The UV terms in Eq. \eqref{eq:Yukawa_neutrino} connect $\xi^{i}_1$ with $\xi^{j}_2$ and are responsible for the neutrino Yukawa coupling. It is an interesting new feature due to the bulk Higgs introduced in the present work, implying that the Dirac neutrino masses can only come from the IR terms if the Higgs field is confined on the IR brane. With the vacuum in Eq.\eqref{eq:vacuum}, the Dirac mass matrix for the zero mode neutrinos is given by
\begin{equation}
\label{eq:mnu}m_{\nu}=\left(\begin{array}{ccc}
a   &  b   &  -\sqrt{2}\;b  \\
-b  &  a   &  0  \\
\sqrt{2}\;b   &  0    &   a
\end{array}\right)\frac{v}{\sqrt{2}}\,,
\end{equation}
where
\begin{equation}
\label{eq:a&b}a=\frac{v_{\zeta}}{\Lambda}\frac{y_1}{\left(\Lambda L\right)^{3/2}}F(y=0,\beta,c_{l},c_{\nu})\,,\quad b=\frac{v_{\phi}}{\Lambda}\frac{y_2}{\left(\Lambda L\right)^{3/2}}F(y=0,\beta,c_{l},c_{\nu})\,.
\end{equation}
In this context, the tiny neutrino masses are dominantly determined by the wavefunction overlaps of the Higgs and the UV scalar $\zeta$ or $\phi$. Since the Higgs profile is localized close to the IR brane, the neutrino masses are highly suppressed. The suppression factor $F(y=0,\alpha,c_{\ell},c_{\nu})$ is given by
\begin{eqnarray}
\hskip-0.14in F(y=0,\beta,c_{l},c_{\nu})=(kL)^{3/2}\sqrt{\frac{2\left(1-\beta\right) \left(1-2c_{l}\right)\left(1+2c_{\nu}\right)} {\left[1-e^{-2(1-\beta)kL}\right]
\left[e^{(1-2c_{l})kL}-1\right]\left[e^{(1+2c_{\nu})kL}-1\right]}}\; e^{-(1-\beta)kL}.
\end{eqnarray}
Therefore the light neutrino masses are roughly suppressed by the factor $e^{-(1-\beta)kL}$ with respect to the electroweak scale, and the correct scale of the neutrino masses could be easily obtained by varying the Higgs localization parameter $\beta$, as will be investigated in detail in the end of this section~\footnote{Note that if we wanted to choose the second solution for the Higgs profile which is proportional to $e^{(2+\beta)k(y-L)}$, all we would have to do is to flip the sign of $\beta$ in all the equations. In other words, taking $\beta<0$ formally corresponds to the other solution. Then the neutrino mass scale would be suppressed by $e^{-(1+\beta)kL}$ relative to the electroweak scale, generally the resulting neutrino masses would be too small to accommodate the measured solar and atmospheric mass-squared splittings without strong fine-tuning.}. Now we come to the diagonalization of the neutrino mass matrix in Eq. \eqref{eq:mnu}, which can be diagonalized by bi-unitary transformations, i.e.,
\begin{equation}
U^{\dagger}_Lm_{\nu}U_R=\text{diag}(m_1,m_2,m_3)\,,
\end{equation}
where the unitary matrices $U_L$ and $U_R$ are
\begin{equation}
\label{eq:UL_UR}U_L=U_R=\left(\begin{array}{ccc}
\frac{i}{\sqrt{2}}  &   -\frac{i}{\sqrt{2}}  &  0  \\
-\frac{1}{\sqrt{6}}  &  -\frac{1}{\sqrt{6}}   &   \sqrt{\frac{2}{3}}  \\
\frac{1}{\sqrt{3}}   &   \frac{1}{\sqrt{3}}   &  \frac{1}{\sqrt{3}}
\end{array}\right)\,.
\end{equation}
It is exactly the DC mixing matrix. Since the charged lepton mass matrix is diagonal, the lepton mixing entirely comes from the neutrino sector, it is of the DC form. The light neutrino masses $m_{1,2,3}$ are given by~\footnote{As the order of the neutrino mass is less constrained experimentally so far, we can also reorder the light neutrino masses such that $m_1=\left(a-i\sqrt{3}\,b\right)v/\sqrt{2}$, $m_2=\left(a+i\sqrt{3}\,b\right)v/\sqrt{2}$ and $m_3=av/\sqrt{2}$. The associated unitary transformations can be obtained from $U_L$ and $U_R$ by permuting its first and second columns. The resulting lepton mixing matrix is still of the DC form, and the sign difference with respect to the DC mixing matrix of Eq. \eqref{eq:UDC} in the first row can be eliminated by redefining the lepton fields. In general, if the LO mixing ansatz gives rise to $\theta_{12}=45^{\circ}$, the lepton flavor mixing is invariant under the exchange of the first and the second light neutrino mass eigenstates. The bimaximal mixing is another typical example.}
\begin{equation}
\label{eq:neutrino_masses}m_1=\left(a+i\sqrt{3}\,b\right)\frac{v}{\sqrt{2}}\;,\quad m_2=\left(a-i\sqrt{3}\,b\right)\frac{v}{\sqrt{2}}\;,\quad m_3=a\frac{v}{\sqrt{2}}\,.
\end{equation}
Obviously the light neutrino masses satisfy the sum rule
\begin{equation}
m_1+m_2=2m_3\,,
\end{equation}
where $m_i$ ($i=1,2,3$) are generally complex numbers, so this relation cannot be used to exactly predict one physical neutrino mass in terms of the other two. Note that this sum rule is different from previous neutrino mass sum rules obtained in the context of Majorana neutrinos \cite{Barry:2010yk}.

\subsection{Phenomenological implications}

\begin{figure}[t!]
\begin{center}
\includegraphics[width=0.85\textwidth]{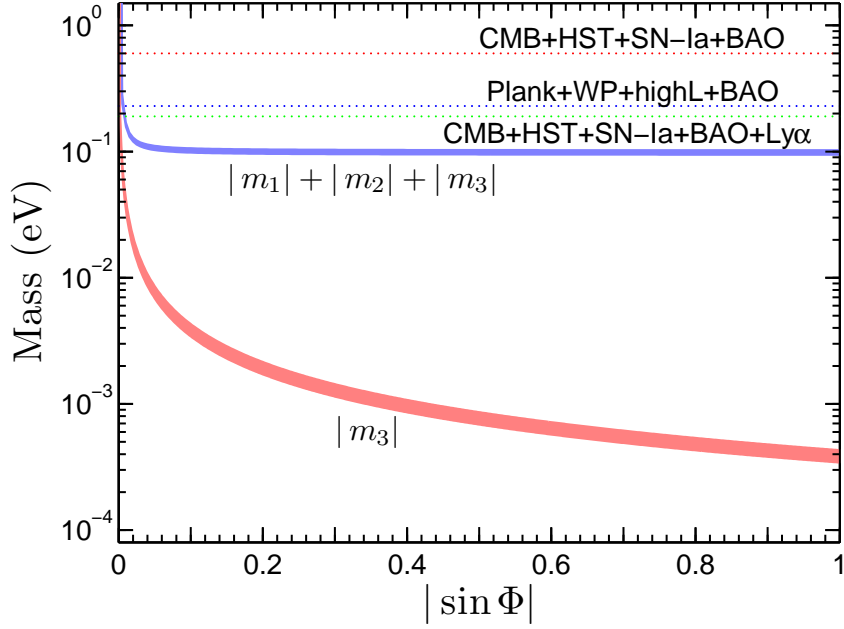}
\end{center}
\caption{\label{fig:neutrino_mass}The lightest neutrino mass $|m_3|$ and the sum of neutrino masses $\sum^{3}_{i=1}|m_i|$ versus $|\sin\Phi|$. The bands have been obtained by varying both $\Delta m^2_{sol}$ and $\Delta m^2_{atm}$ in their $3\sigma$ experimental ranges.}
\end{figure}

As shown above, the neutrino part of the model depends on two complex parameters $a$ and $b$ at LO. Absorbing the overall phase by field redefinition, we are left with three real parameters, which can be chosen to be $|a|$, $|b|$ and $\Phi=\text{arg}(b)-\text{arg}(a)$ with $\Phi$ being the relative phase between the complex numbers $a$ and $b$. Now we turn to its predictions for the neutrino mass spectrum. From Eq. \eqref{eq:neutrino_masses} we have
\begin{equation}
|m_1|^2,|m_2|^2=\left(|a|^2+3|b|^2\mp2\sqrt{3}\;|a||b|\sin\Phi\right)\frac{v^2}{2}\,,\\
\quad
|m_3|^2=|a|^2\frac{v^2}{2}\,,
\label{eq:mi}
\end{equation}
which leads to~\footnote{In this case, one can easily deduce $\sin\Phi>0$ from the constraint $\Delta m_{sol}^2\equiv|m_2|^2-|m_1|^2=2\sqrt{3}\;v^2|a||b|\sin\Phi>0$, while one has $\sin\Phi<0$ for the other mass ordering $m_1=\left(a-i\sqrt{3}\,b\right)v/\sqrt{2}$ and $m_2=\left(a+i\sqrt{3}\,b\right)v/\sqrt{2}$. }
\begin{equation}
|m_2|^2>(|m_1|^2+|m_2|^2)/2>|m_3|^2 \,,
\end{equation}
where we have employed the constraint $|m_2|^2>|m_1|^2$ from the solar neutrino oscillation experiments in the first inequality. Therefore the light neutrino mass spectrum can only be inverted hierarchy in the present model. Experimentally, the solar and atmospheric mass-squared splittings have been measured precisely. For the inverted hierarchy, they are \cite{GonzalezGarcia:2012sz}
\begin{eqnarray}
\nonumber&&\Delta m^2_{sol}\equiv|m_2|^2-|m_1|^2=(7.50^{+0.18}_{-0.19})\times10^{-5} \mathrm{eV}^2,\\
&&\Delta m^2_{atm}\equiv|m_2|^2-|m_3|^2=\left(2.427^{+0.065}_{-0.042}\right)\times10^{-3} \mathrm{eV}^2\,.
\end{eqnarray}
We can express the three light neutrino masses in terms of $r\equiv\Delta m^2_{sol}/\Delta m^2_{atm}$, $\Delta m^2_{atm}$ and $\sin\Phi$ as
\begin{eqnarray}
\nonumber&&|m_1|^2=\left[1-r+\frac{r^2}{8(2-r)\sin^2\Phi}\right]\Delta m^2_{atm}\,,\\
\nonumber&&|m_2|^2=\left[1+\frac{r^2}{8(2-r)\sin^2\Phi}\right]\Delta m^2_{atm}\,,\\
&&|m_3|^2=\frac{r^2}{8(2-r)\sin^2\Phi}\;\Delta m^2_{atm}\,.
\end{eqnarray}
Note that these relations are also satisfied even if we switch the first and second light neutrino masses shown in Eq. \eqref{eq:neutrino_masses}. Thus we obtain the following limit for the lightest neutrino mass $m_3$,
\begin{equation}
|m_3|\geq3.8\times10^{-4}\; \text{eV}\,,
\end{equation}
where the lower bound corresponds to $|\sin\Phi|=1$. This implies that the light neutrino mass spectrum has a partial hierarchy:
\begin{equation}
|m_1|\simeq4.8\times10^{-2}\;\text{eV}\,,\qquad |m_2|\simeq4.9\times10^{-2}\;\text{eV}\,,\qquad |m_3|\simeq3.8\times10^{-4}\; \text{eV}\,,
\end{equation}
and the sum of neutrino masses is about $9.8\times10^{-2}\;\text{eV}$. If $\sin\Phi$ is accidentally small, the neutrino spectrum becomes degenerate. We have plotted the neutrino masses
predicted by the model in Fig. \ref{fig:neutrino_mass}, where the horizontal lines are the cosmological bounds \cite{Fogli:2008cx}. The first one at $0.60$ eV corresponds to the combination of the Cosmic Microwave Background
(CMB) anisotropy data (from WMAP~5y \cite{WMAP2}, Arcminute
Cosmology Bolometer Array Receiver (ACBAR) \cite{acbar07}, Very
Small Array (VSA) \cite {vsa}, Cosmic Background Imager (CBI)
\cite{cbi} and BOOMERANG \cite{boom03} experiments) plus the
large-scale structure (LSS) information on galaxy clustering (from
the Luminous Red Galaxies Sloan Digital Sky Survey (SDSS)
\cite{Tegmark}) plus the Hubble Space Telescope (HST) plus the
luminosity distance SN-Ia data of \cite{astier} and finally plus the
BAO data from \cite{bao}. The second one at $0.19$ eV corresponds to
all the previous data combined to the small scale primordial
spectrum from Lyman-alpha (Ly$\alpha$) forest clouds \cite{Ly1}. Furthermore, the most recent Planck results $\sum_{i}m_i<0.23~\text{eV}$ are also shown~\cite{Ade:2013zuv}. We see that the sum of light neutrino masses in the present model is below the present cosmological bound except for extremely small $\sin\Phi$.

In the following, we give an illustrative example, a set of
parameters reproducing the charged lepton masses and the neutrino mass-squared differences. We take the fundamental 5D scale to be $k\simeq\Lambda\simeq M_{Pl}$, where $M_{Pl}\simeq2.44\times10^{18}$ GeV is the reduced Planck mass. We set the value $k'=ke^{-kL}\simeq 1.5$ TeV in order to resolve the hierarchy between the Planck and electroweak symmetry breaking scales, and the resulting first KK gauge bosons with mass $m_{KK}=3\sim4$ TeV are in the reach of LHC experiments. The Higgs VEV turns out to be $v\simeq250$ GeV. It is obtained by matching the precisely measured fine-structure constant, Fermi constant and the Z boson mass with the corresponding ones of the present warped model, as will be shown in section \ref{sec:EWC}, The parameter $V/\Lambda$ is fixed at the indicative value of 0.15 which is implied by the higher order corrections to achieve agreement with the measured lepton mixing angles, where $V$ stands for a generic flavon VEV ($v_{\chi}$, $v_{\varphi}$, $v_{\zeta}$, $v_{\phi}$ or $v_{\rho}$). The Higgs localization parameter $\beta$ will be set to 0.25 for demonstration in the following.

\begin{figure}[t!]
\begin{center}
\includegraphics[width=0.85\textwidth]{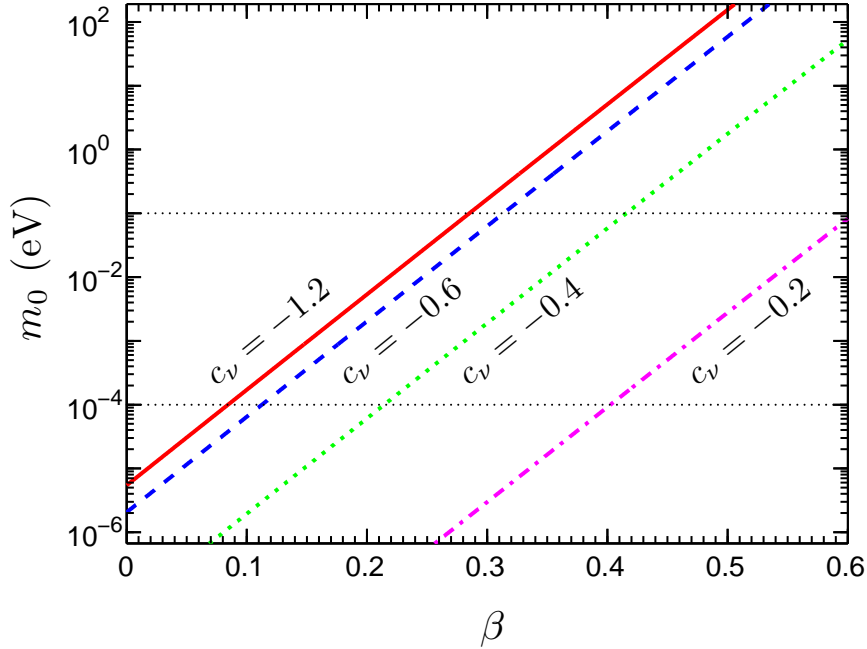}
\caption{\label{fig:neutrino_mass_scale} The light neutrino mass scale $m_0$ versus the bulk Higgs parameter $\beta$ for typical UV localized RH neutrino $c_{\nu}=-1.2$, $c_{\nu}=-0.6$ and IR localized RH neutrino $c_{\nu}=-0.4$, $c_{\nu}=-0.2$, where $c_l$ is taken to be 0.51. The horizontal lines correspond to $m_0$ being $10^{-4}$ eV and $10^{-1}$ eV, respectively.}
\end{center}
\end{figure}

In the charged lepton sector, we make the choice $c_{l}=0.51$, $c_{e}=-0.70$, $c_{\mu}=-0.58$, $c_{\tau}=-0.502$, $y_{e}=1.248$, $y_{\mu}=0.932$ and $y_{\tau}=2.363$. These values give the charged lepton masses $m_{e}\simeq0.490$ MeV, $m_{\mu}\simeq103.3$ MeV and $m_{\tau}\simeq1.757$ GeV, which are consistent with the corresponding values around 1 TeV scale~\cite{Xing:2007fb}. In the neutrino sector, clearly the light neutrino mass scale is characterized by $m_0\equiv F(y=0,\beta,c_l,c_{\nu})Vv/(\sqrt{2}\Lambda)$ . For the Higgs profile of Eq. \eqref{eq:higgs_profile}, $m_0$ increases with the Higgs bulk mass parameter $\beta$ sharply, as is displayed in Fig. \ref{fig:neutrino_mass_scale}. From this figure, we can see that the curve moves from the upper left corner to the lower right corner with the increase of $c_{\nu}$. If the RH neutrino is localized close to the UV brane ($c_{\nu}<-0.5$), $m_0$ weakly depends on $c_\nu$. Taking for example $c_{\nu}=-0.6$, the neutrino mass in the range of $10^{-4}\sim10^{-1}$ eV can be obtained for $\beta$ in the interval $0.11<\beta<0.31$. We note that this $\beta$ range varies slightly with the UV type $c_{\nu}$. On the other hand, if the RH neutrino is localized near the IR brane ($c_{\nu}>-0.5$), the localization parameter $\beta$ should be larger to account for the correct magnitude of neutrino masses, since they are approximately suppressed by the factor $e^{-\left(3/2-\beta+c_{\nu}\right)kL}$ in this case~\footnote{If we had selected the other solution for the Higgs profile which is proportional to $e^{(2+\beta)k(y-L)}$, $m_0$ should have decreased with increasing $\beta$, and $m_0$ should have been typically of order $10^{-6}\sim10^{-5}$ eV for small $\beta\sim0$ and $-c_{\nu}\sim\mathcal{O}(1)$. To produce the correct order of magnitude of the neutrino masses, the couplings $y_1$ and $y_2$ would be very large, about several thousands, or the RH neutrinos would be localized extremely close to UV brane, such that their bulk mass parameters should be around tens of thousands.
This is exactly the reason why we choose the Higgs profile of Eq. \eqref{eq:higgs_profile}.}. We use the value of $c_l$ above, $c_{\nu}=-0.6$, $y_1=0.95$ and $y_2=2.49+0.09i$. These parameters give absolute neutrino masses $m_1\simeq0.0496$ eV, $m_2\simeq0.0503$ eV and $m_3\simeq0.0107$ eV. These correspond to the mass squared differences $\Delta m^2_{sol}\simeq7.57\times10^{-5}\;\mathrm{eV}^2$ and $\Delta m^2_{atm}\simeq2.418\times10^{-3}\;\mathrm{eV}^2$, which are in good agreement with the experimental values \cite{Tortola:2012te,Fogli:2012ua,GonzalezGarcia:2012sz}.

\section{\label{sec:NLO} Higher order corrections}

The results of the previous section hold to the LO approximation. High dimensional operators, suppressed by additional powers of the cut-off $\Lambda$, can be added to the leading terms in Eqs. \eqref{eq:Yukawa_charged} and \eqref{eq:Yukawa_neutrino}. Another motivation for considering higher order corrections is to obtain an acceptable lepton mixing pattern in particular to get an appreciable $\theta_{13}$, since the LO exact DC mixing implies three mixing angles $\sin^2\theta^{DC}_{12}=1/2$, $\sin^2\theta^{DC}_{23}=2/3$ and $\sin^2\theta^{DC}_{13}=0$, which are sizeably different from the experimentally measured values.

We first focus on the IR brane. The most general higher order operators invariant under the imposed symmetry take the following form
\begin{eqnarray}
\nonumber\delta \mathcal{L}_{IR}&=&\sum_{m\geq2,n\geq0}\frac{1}{\Lambda^{(m+2n+3/2)}}\big[\text{Tr}\left(\overline{\xi}_1T^{3}_3H\right)+\text{Tr}\left(\overline{\xi}_1HT^{3T}_4\right)\big]\left(\chi^m\varphi^{2n}\right)_{\mathbf{3}}\\
\nonumber&+&\sum_{m\geq0,n\geq1}\frac{1}{\Lambda^{(m+2n+5/2)}}\big[\text{Tr}\left(\overline{\xi}_1T^{2}_3H\right)+\text{Tr}\left(\overline{\xi}_1HT^{2T}_4\right)\big]\left(\chi^{m}\varphi^{2n+1}\right)_{\mathbf{3}'}\\
\label{eq:corrections_charged}&+&\sum_{m\geq1,n\geq1}\frac{1}{\Lambda^{(m+2n+5/2)}}\big[\text{Tr}\left(\overline{\xi}_1T^{1}_3H\right)+\text{Tr}\left(\overline{\xi}_1HT^{1T}_4\right)\big]\left(\chi^m\varphi^{2n+1}\right)_{\mathbf{3}}+\text{h.c.}\,,
\end{eqnarray}
where the overall factor $-2\sqrt{G}\;\delta(y-L)$ is omitted, and we have suppressed the order one coupling constant in front of each operator. The contributions of the above operators can be understood easily from symmetry. From the Appendix A, in the triplet representations, the representation matrix for each element of the Klein four subgroup $K_4=\left\{1,G_1=TSTS^2,\;G_2=TST,\;G_3=S^2\right\}$ is
\begin{equation}
G_1=\pm\left(\begin{array}{ccc}
-1 &  0 & 0  \\
0  & 1  &  0  \\
0  & 0  & 1
\end{array}\right),\quad G_2=\pm\left(\begin{array}{ccc}
1  &   0   &   0  \\
0  &  -1   &   0   \\
0  &  0    &    1
\end{array}\right),\quad G_3=\left(\begin{array}{ccc}
-1  &   0   &  0  \\
0   &   -1  &   0  \\
0   &   0   &   1
\end{array}\right)\,,
\end{equation}
where the sign $``+"$ for the representation $\mathbf{3}$ and $``-"$ for $\mathbf{3}'$. It is straightforward to check that
\begin{eqnarray}
\nonumber&G_1\langle\chi\rangle=G_2\langle\chi\rangle=G_3\langle\chi\rangle=\langle\chi\rangle\,,\\
&-G_1\langle\varphi\rangle=G_2\langle\varphi\rangle=-G_3\langle\varphi\rangle=\langle\varphi\rangle\,.
\end{eqnarray}
As a result, we have
\begin{eqnarray}
\nonumber&G_1\left\langle\left(\chi^m\varphi^{2n}\right)_{\mathbf{3}}\right\rangle=G_2\left\langle\left(\chi^m\varphi^{2n}\right)_{\mathbf{3}}\right\rangle=G_3\left\langle\left(\chi^m\varphi^{2n}\right)_{\mathbf{3}}\right\rangle=\left\langle\left(\chi^m\varphi^{2n}\right)_{\mathbf{3}}\right\rangle\,,\\
\nonumber&-G_1\big\langle\left(\chi^{m}\varphi^{2n+1}\right)_{\mathbf{3}\left(\mathbf{3}'\right)}\big\rangle=G_2\big\langle\left(\chi^{m}\varphi^{2n+1}\right)_{\mathbf{3}\left(\mathbf{3}'\right)}\big\rangle=-G_3\big\langle\left(\chi^{m}\varphi^{2n+1}\right)_{\mathbf{3}\left(\mathbf{3}'\right)}\big\rangle
=\big\langle\left(\chi^{m}\varphi^{2n+1}\right)_{\mathbf{3}\left(\mathbf{3}'\right)}\big\rangle\,.
\end{eqnarray}
Thus, the vacuum alignments of $\left(\chi^m\varphi^{2n}\right)_{\mathbf{3}}$ and the combination $\left(\chi^{m}\varphi^{2n+1}\right)_{\mathbf{3},\mathbf{3}'}$ are
\begin{equation}
\left\langle\left(\chi^m\varphi^{2n}\right)_{\mathbf{3}}\right\rangle\propto(0,0,1)^{T},~~ \big\langle\left(\chi^{m}\varphi^{2n+1}\right)_{\mathbf{3}'}\big\rangle\propto(0,1,0)^{T},~~ \big\langle\left(\chi^{m}\varphi^{2n+1}\right)_{\mathbf{3}}\big\rangle\propto(1,0,0)^{T}.
\end{equation}
They are along the same directions as the VEVs $\langle\chi\rangle$, $\langle\varphi\rangle$ and $\left\langle\left(\chi\varphi\right)_{\mathbf{3}}\right\rangle$, respectively. As a result, the high dimensional terms in Eq. \eqref{eq:corrections_charged} only lead to corrections to the electron, muon and tau masses, all the off-diagonal entries of the charged lepton mass matrix are still vanishing, and their contributions can be absorbed by a redefinition of the LO parameters $y_e$, $y_{\mu}$ and $y_{\tau}$. This means that the charged lepton mass matrix is diagonal even when higher order operators are taken into account, and no rotation of the charged lepton fields to the mass eigenstates is needed.

On the UV brane, the NLO corrections to the Dirac neutrino Lagrangian of Eq. \eqref{eq:Yukawa_neutrino} can be written as
\begin{eqnarray}
\hskip-0.14in
\delta\mathcal{L}_{UV}=\frac{z_1}{\Lambda^{7/2}}\text{Tr}\left(\overline{\xi}_1H\xi_2\right)\zeta\rho+\frac{z_2}{\Lambda^{7/2}}\text{Tr}\left(\overline{\xi}_1H\xi_2\right)\left(\rho\phi\right)_{\mathbf{3}}
+\frac{z_3}{\Lambda^{7/2}}\text{Tr}\left(\overline{\xi}_1H\xi_2\right)\left(\rho\phi\right)_{\mathbf{3}'}+\text{h.c.}\,,
\end{eqnarray}
where the overall factor $\sqrt{2G}\;\delta(y)$ is neglected. Expanding the above terms according to the contraction rules listed in the Appendix A, and after the $S_4$ flavor symmetry breaking by the VEVs in Eq. \eqref{eq:vacuum}, we obtain the corrections to the neutrino mass matrix
\begin{equation}
\label{eq:mnu_nlo}\delta m_{\nu}=\left(\begin{array}{ccc}
\epsilon_1   &  -\sqrt{2}\;\epsilon_3   &  \epsilon_2-\epsilon_3  \\
\sqrt{2}\;\epsilon_3   &  \epsilon_1   &  0  \\
\epsilon_2+\epsilon_3   &  0   &  -2\epsilon_1
\end{array}\right)\frac{v}{\sqrt{2}}\,,
\end{equation}
where the parameters $\epsilon_{1,2,3}$ are given by
\begin{eqnarray}
\nonumber&&\epsilon_1=\frac{z_1}{\left(\Lambda L\right)^{3/2}}\frac{v_{\zeta}v_{\rho}}{\Lambda^2}\;F\left(y=0,\beta,c_{l},c_{\nu}\right),\\
\nonumber&&\epsilon_2=\sqrt{6}\frac{z_2}{\left(\Lambda L\right)^{3/2}}\frac{v_{\rho}v_{\phi}}{\Lambda^2}\;F\left(y=0,\beta,c_{l},c_{\nu}\right),\\ &&\epsilon_3=\sqrt{2}\frac{z_3}{\left(\Lambda L\right)^{3/2}}\frac{v_{\rho}v_{\phi}}{\Lambda^2}\;F\left(y=0,\beta,c_{l},c_{\nu}\right),
\end{eqnarray}
which are suppressed by $v_{\rho}/\Lambda$ with respect to the LO light neutrino parameters $a$ and $b$ of Eq. \eqref{eq:a&b}. To first order in the small parameters $\epsilon_i (i=1,2,3)$, the lepton mixing angles are modified to
\begin{eqnarray}
\nonumber\sin\theta_{13}&=&\frac{1}{\sqrt{3}\;\left[3|b|^4+(a\bar{b}-\bar{a}b)^2\right]}\Big|\sqrt{2}\;\bar{a}\left(a\bar{b}-\bar{a}b\right)\epsilon_1+\bar{a}|b|^2\left(\epsilon_2-3\epsilon_3\right)\,,
\\
\nonumber&&+\sqrt{2}\;\left[(a^2+3b^2)\bar{b}-|a|^2b\right]\bar{\epsilon}_1+\bar{a}b^2\left(\bar{\epsilon}_2+3\bar{\epsilon}_3\right)
\Big|\,,\\
\nonumber\sin^2\theta_{12}&=&\frac{1}{2}+\frac{1}{2\sqrt{3}\;i\left(a\bar{b}-\bar{a}b\right)}\Big\{\bar{a}\epsilon_1-\sqrt{2}\;\bar{b}\epsilon_2+c.c.\Big\}\,,\\
\nonumber\sin^2\theta_{23}&=&\frac{2}{3}-\frac{\sqrt{2}}{3\left[3|b|^4+(a\bar{b}-\bar{a}b)^2\right]}\Big\{\sqrt{2}\;\left(a\bar{b}^2+\bar{a}|b|^2\right)\epsilon_1
-|b|^2\bar{b}\epsilon_2\\
&&-\left[\left(2\bar{a}^2+3\bar{b}^2\right)b-2|a|^2\bar{b}\right]\epsilon_3+c.c.\Big\}\,,
\end{eqnarray}
where a bar on a letter indicates complex conjugation. It is clear that the NLO corrections to the three mixing angles are of order $v_{\rho}/\Lambda$. We see that, in order to be compatible with the current experimental values of the mixing angles \cite{Tortola:2012te,Fogli:2012ua,GonzalezGarcia:2012sz}, we need a parameter $v_{\rho}/\Lambda$ of the order $0.1\sim0.2$, which is approximately the size of the Cabibbo angle $\lambda_c$. In particular, the resulting reactor mixing angle $\theta_{13}$ would be expected to be of the same order if no accidental cancellations among the coefficients occur. Regarding the light neutrino masses, these corrections introduce deviations in $\epsilon^2_i\;(i=1,2,3)$ with respect to the LO predictions of Eq. \eqref{eq:neutrino_masses}, and therefore the neutrino spectrum should also be an inverted hierarchy even after the high dimensional terms are taken into account.

We see that the NLO contributions introduce three additional complex parameters $z_1$, $z_2$ and $z_3$ in the neutrino sector. Taking into account the LO parameters $y_1$, $y_2$ and the bulk mass parameters $c_{l}$, $c_{\nu}$ and $\beta$, we have enough independent parameters to fit the neutrino masses and mixing angles. For example, we set the representative values $y_1=-1.572-0.654 i$, $y_2=1.785+0.777 i$, $z_1=0.571-0.281 i$, $z_2=0.359-1.270 i$, $z_3=-0.199-1.986 i$ and $\beta=0.2564$, and we take $c_l=0.51$, $c_{\nu}=-0.60$ and $V/\Lambda=0.15$ following the LO case, where $V$ denotes the VEV of any flavon field. Then the lepton mixing parameters and the light neutrino masses are determined to be $\sin^2\theta_{12}\simeq0.302$, $\sin^2\theta_{23}\simeq0.593$, $\sin^2\theta_{13}\simeq0.0232$, $m_1\simeq0.0545\;\text{eV}$, $m_2\simeq0.0552\;\text{eV}$ and $m_3\simeq0.0231\;\text{eV}$, which give rise to $\Delta m^2_{sol}\simeq7.679\times10^{-5}\;\text{eV}^2$ and $\Delta m^2_{atm}\simeq2.513\times10^{-3}\;\text{eV}^2$. These predictions are in excellent agreement with the present data~\cite{Tortola:2012te,Fogli:2012ua,GonzalezGarcia:2012sz}.

\begin{figure}[t!]
\begin{center}
\includegraphics[width=\textwidth]{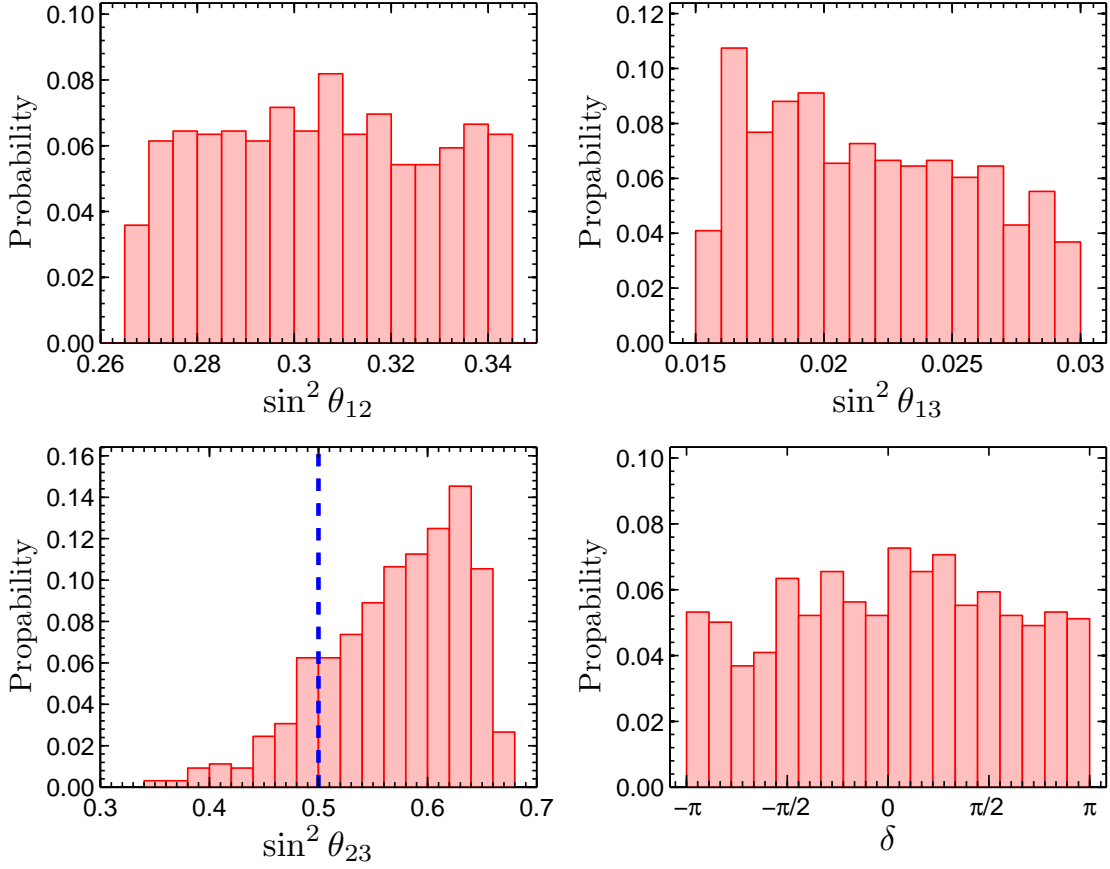}
\caption{\label{fig:histogram_DC}The probability distribution of the lepton mixing parameters $\sin^2\theta_{12}$, $\sin^2\theta_{13}$, $\sin^2\theta_{23}$ and the Dirac CP phase $\delta$ in the DC model. }
\end{center}
\end{figure}

In order to see in a quantitative way the behavior of the lepton mixing parameters under the NLO corrections, we perform a numerical analysis in the following. The parameter $V/\Lambda$ has been fixed at the representative value of 0.15. where $V$ stands for the VEV of any flavon field in the model. All the undetermined order one dimensionless couplings are randomly generated complex numbers, the absolute values are taken to be uniformly distributed in the interval $\left[1/2,2\right]$, and the phases are chosen to be flatly distributed in $\left[0,2\pi\right)$. Then we calculate the associated neutrino masses and the mixing angles which are required to lie in the $3\sigma$ ranges \cite{Tortola:2012te,Fogli:2012ua,GonzalezGarcia:2012sz}. With the procedure described above, we find that the neutrino mass spectrum is inverted hierarchy for all the generated points, which is consistent with the conclusion reached from the above analytical estimates. Determination of the neutrino mass hierarchy is one of the primary physics goals of the next generation precise neutrino oscillation experiments. If the mass spectrum is determined to be normal hierarchy in future, our construction would be ruled out. As is generally true in flavor symmetry models, any quantitative estimates are affected by large uncertainties due to the presence of unknown parameters of order one, if the NLO contributions are taken into account. The distributions of the three mixing angles and the CP phase $\delta$ are plotted in Fig. \ref{fig:histogram_DC}. From this figure, we can hardly see any specifically preferred pattern for $\theta_{12}$, and smaller $\theta_{13}$ is slightly favored as $\theta_{13}=0$ at LO. The distribution of $\delta$ is essentially random, since our model doesn't introduce any mechanisms to predict certain specific values. However, the atmospheric neutrino mixing angle $\theta_{23}$ with $\sin^2\theta_{23}>1/2$, which means the second octant of $\theta_{23}$, is clearly preferred. This fact can be easily understood as follows. The atmospheric mixing angle is predicted to be the DC value $\sin^2\theta^{DC}_{23}=2/3>1/2$ at LO, after including the NLO contributions, the first octant $\theta_{23}$ could only be achieved if the NLO corrections are rather large. Nevertheless, this requires an additional fine-tuning of the parameters which has been reproduced in our numerical simulation only partially and by a few points. At present, significant deviation of $\theta_{23}$ from maximal mixing value $45^{\circ}$ is favored at the level of $1.7-2\; \sigma$ \cite{Tortola:2012te,Fogli:2012ua,GonzalezGarcia:2012sz}, and we still don't know which octant $\theta_{23}$ lies in. However, future long-baseline neutrino oscillation experiments open up the possibility of determining the octant of $\theta_{23}$ \cite{Feldman:2012qt}. If future experiments discover that $\theta_{23}$ belongs to the second octant and the deviation from maximal mixing is somewhat large, the present suggestion of DC mixing as a starting mixing pattern of model construction would be strongly favored.

\begin{figure}[t!]
\begin{center}
\includegraphics[width=0.85\textwidth]{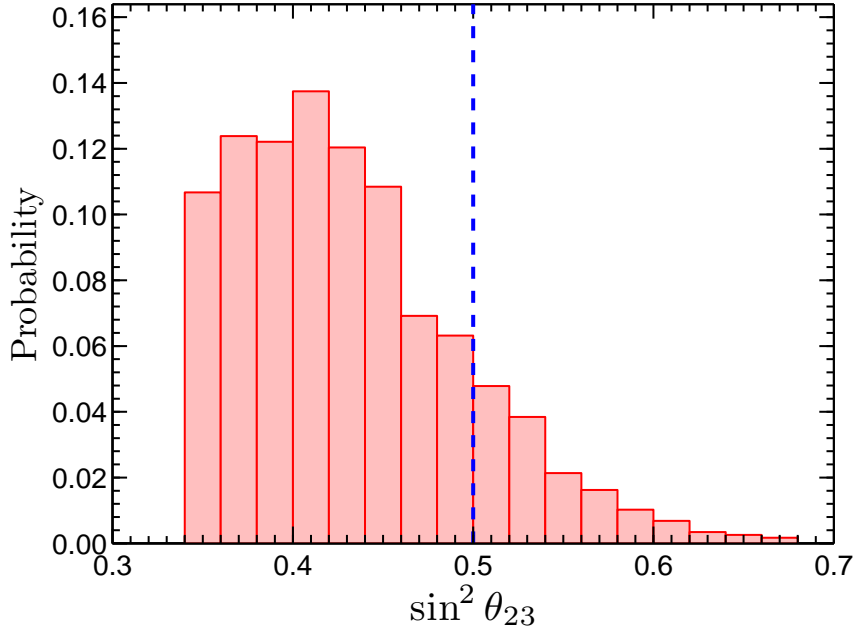}
\end{center}
\caption{\label{fig:histogram_MDC}The distribution of the atmospheric mixing angle $\sin^2\theta_{23}$, where the LO mixing is the MDC texture.}
\end{figure}

The deviation of $\theta_{23}$ from maximality, if confirmed in future, would have profound implications for neutrino mass models based on flavor symmetries. Since the DC mixing is preferred for the second octant of $\theta_{23}$, we would like to propose another new texture which could be taken as a LO approximation for the first octant of $\theta_{23}$.  By permuting the second and third rows of the DC mixing matrix, we find a new interesting mixing pattern which is called ``modified" DC (MDC) mixing with
\begin{equation}
U_{MDC}=\left(\begin{array}{ccc}
\frac{i}{\sqrt{2}}  &  -\frac{i}{\sqrt{2}}  &  0\\
\frac{1}{\sqrt{3}}  &  \frac{1}{\sqrt{3}}   &  \frac{1}{\sqrt{3}}  \\
-\frac{1}{\sqrt{6}}  &  -\frac{1}{\sqrt{6}} &  \sqrt{\frac{2}{3}}
\end{array}\right)\,.
\end{equation}
It gives rise to the lepton mixing angles
\begin{equation}
\sin^2\theta^{MDC}_{13}=0,\qquad \sin^2\theta^{MDC}_{12}=1/2,\qquad  \sin^2\theta^{MDC}_{23}=1/3\,.
\end{equation}
Obviously we have $\theta_{23}^{MDC}=\pi/2-\theta_{23}^{DC}$. Similar to the DC mixing case, the MDC pattern should be corrected by terms of $\mathcal{O}(\lambda_c)$ in order to achieve agreement with the experimental data. It is easy to modify the DC model of section \ref{sec:model} to get an MDC model by embedding $\left(\xi^{1}_1,\xi^{3}_1,\xi^{2}_1\right)$ instead of $\left(\xi^{1}_1,\xi^{2}_1,\xi^{3}_1\right)$ into an $S_4$ triplet $\mathbf{3}$ and by exchanging the assignments for $\xi^{2}_3$ and $\xi^{3}_3$ under the flavor symmetry. Then we can straightforwardly check that the charged lepton mass matrix is also diagonal even in the presence of the high dimensional operators. The LO and NLO neutrino mass matrices $m'_{\nu}$ and $\delta m'_{\nu}$ can be obtained from $m_{\nu}$ and $\delta m_{\nu}$ respectively by switching the second and the third rows
\begin{equation}
m'_{\nu}=P_{23}m_{\nu},\qquad \delta m'_{\nu}=P_{23}\delta m_{\nu}\,,
\end{equation}
where $m_{\nu}$ and $\delta m_{\nu}$ in Eq. \eqref{eq:mnu} and Eq. \eqref{eq:mnu_nlo} are the predictions of the above DC model for the LO and NLO neutrino mass matrices respectively, and $P_{23}$ is a permutation matrix
\begin{equation}
P_{23}=\left(\begin{array}{ccc}
1  & 0  & 0  \\
0  & 0  &  1  \\
0  & 1  &   0
\end{array}\right)\,.
\end{equation}
Note that $P_{23}$ arises due to the exchange of the order of $\xi^{2}_1$ and $\xi^{3}_1$ within the triplet $\mathbf{3}$. As a result, the lepton flavor mixing matrix is of the MDC form at LO, the predictions for the neutrino masses are given in Eq. \eqref{eq:neutrino_masses}, and thus the neutrino mass spectrum is inverted hierarchy as well. Following the method outlined in the DC case, we perform a numerical analysis to account for the NLO corrections, the inverted hierarchy neutrino spectrum turns out to be still preserved. The predictions for the distribution of the atmospheric mixing angle $\theta_{23}$ is displayed in Fig. \ref{fig:histogram_MDC}, as expected that the first octant of $\theta_{23}$ is obviously favored. Regarding the mixing angles $\theta_{13}$ and $\theta_{12}$ and the Dirac CP phase $\delta$, no specific values are preferred within $3\sigma$, hence not shown in the figure.

\section{\label{sec:EWC}Electroweak constraints}

The electroweak precision measurements (EWPMs) provide stringent constraints to new physics at the TeV scale. It is well-known that the custodial symmetry can relax the bound of the KK scale \cite{Agashe:2003zs} in the warped extra dimension. It has been shown that an $A_4$ lepton model with an IR brane localized Higgs \cite{Csaki:2008qq} can pass the EWPM constraints for the KK scale of order 3 TeV if fermions have proper bulk masses. In the following, we shall extend the analysis to the bulk Higgs case in the warped extra dimension with the custodial symmetry $SU(2)_L\times SU(2)_R\times U(1)_X\times P_{LR}$, and we shall check that our model can indeed pass the EWPM constraints. In addition, we find it is easier for a bulk Higgs to pass the EWPM constraints than a brane Higgs.

\begin{figure}[t!]
\begin{center}
\includegraphics[width=\textwidth] {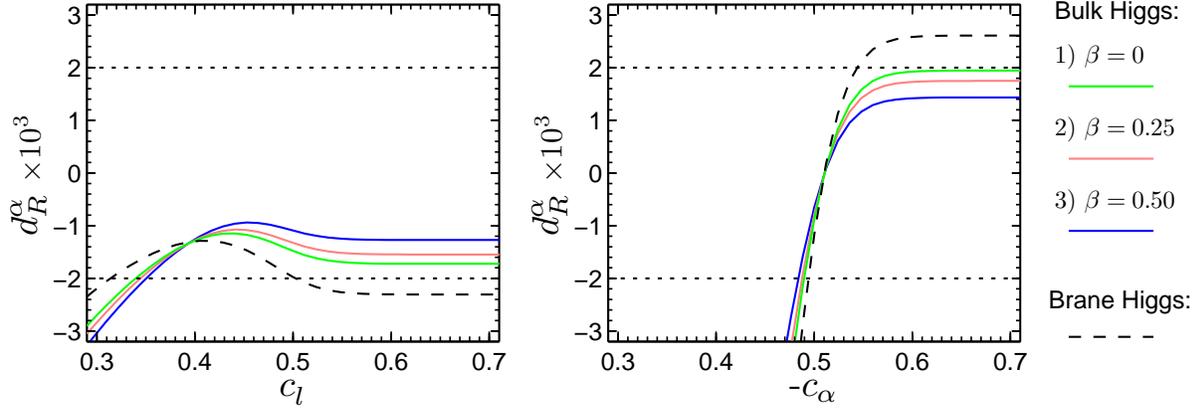}
\caption{\label{d_LR_custodial}Deviations from the SM $Z$ couplings with the LH charged lepton $\ell_{L}$ and the RH charged lepton $\ell_{R}$ in the warped extra dimension with custodial symmetry $SU(2)_L\times SU(2)_R\times U(1)_X\times P_{LR}$. The solid lines are the results for the bulk Higgs scenario considered in the present work, where $\beta$ is taken to be 0 (green), 0.25 (red) and 0.50 (blue), respectively. The result for the brane Higgs scenario is presented in dashed line for comparison. We take $\alpha(\mu=m_Z)=1/128$, $m_Z=91.1876$ GeV, $G_F=1.1664\times10^{-5}~\rm GeV^{-2}$, and $k'=1.5$ TeV as input parameters, and for $d^\alpha_{R}$, we set $c_l=0.51$. The dotted lines stand for the current experimental bounds ($\pm2\times10^{-3}$). }
\end{center}
\end{figure}

We introduce the 4D gauge coupling constants $g$ and $g'$, which are related to the 5D gauge coupling constants via $g=g_5/\sqrt{L}$ and $g'=g^\prime_5/\sqrt{L}$. A simple way of checking the EWPM constraints is to canonically normalize the SM gauge fields and to determine the parameters $g$, $g'$, and $v$ by requiring that the well-measured values of the $Z$ boson mass $m^{}_Z$, the fermi constant $G^{}_F$, and the electrical charge $e$ (or the fine-structure constant $\alpha$) be reproduced. Then the corrections to the EW precision observables would be contained in the gauge couplings of the fermions. We will use the notation that the SM parameters are written with a hat on top of them. The SM expressions for $e$, $G_F$ and $m_Z$ in terms of the $\hat{g}$, $\hat{g}'$, and $\hat{v}$ are given by
\begin{eqnarray}
e=\frac{\hat{g}\hat{g}^{\prime}}{\sqrt{\hat{g}^2+\hat{g}^{\prime2}}}, \hspace{1cm}
m^2_Z=\frac{1}{4}(\hat{g}^2+\hat{g}^{\prime2})\hat{v}^2 \,, \hspace{1cm}
G_F=\frac{1}{\sqrt{2}\;\hat{v}^2} \,.
\label{eq:SM_observable}
\end{eqnarray}
In the warped extra dimension, they are corrected by the massive KK modes, and the main contributions originate from the gauge sector. Following \cite{Carena:2006bn}, we define the zero-momentum gauge boson propagators for the massive KK modes with BCs $(++)$ and $(-+)$, respectively:
\begin{eqnarray}
G_{++}(y,y')&=&\frac{1}{4k(kL)}\left\{\frac{1-e^{2kL}}{kL}+e^{2ky_<}(1-2ky_<)+e^{2ky_>}\left[1+2k(L-y_>)\right]\right\}\,,\nonumber\\
G_{-+}(y,y')&=&-\frac{1}{2k}\left[e^{2ky_<}-1\right]\,,
\end{eqnarray}
where $y_<$ ($y_>$) is the minimum (maximum) of $y$ and $y'$.
To the first order of $v^2/k^{\prime2}$, the parameters $e$, $G_F$, and $m_Z$ in the present warped model are expressed as \cite{Carena:2006bn}
\begin{eqnarray}
&&e=\frac{gg^{\prime}}{\sqrt{g^2+g^{\prime2}}}\,,\nonumber\\
&&m^2_Z=\frac{g^2v^2}{4c^2} \left\{1+\frac{(g^2+g^{\prime2})v^2}{4k^{\prime2}} \left[\delta^{}_{++}+\left(c^2-s^2\right)\delta_{-+}\right]\right\}  \,, \nonumber\\
&&G_F=\frac{1}{\sqrt{2}v^2} \left\{ 1-\frac{g^2v^2}{4k^{\prime2}} \left[\delta_{++}+\delta_{-+}-2G^l_{++}+2G^l_{-+}+G^{\mu\mu}_{++}\right] \right\} \,,\label{eq:RS_observable}
\end{eqnarray}
where $c=g/\sqrt{g^2+g^{\prime2}}$ and $s=g'/\sqrt{g^2+g^{\prime2}}$. The dimensionless parameters $\delta^{}_{\pm+}$ and $G^{l}_{\pm+}$ are corrections induced by the exchange of massive KK gauge bosons with BCs $(\pm+)$, and $G^{\mu\mu}_{++}$ represents additional KK exchange corrections to 4-fermion interactions that contribute to muon decay. They are explicitly given by
\begin{eqnarray}
\delta^{}_{\pm+}&=&\frac{k^{\prime2}}{L} \int^L_0dydy'e^{-2ky}h^2(y)G_{\pm+}(y,y') e^{-2ky'}h^2(y')\,,\nonumber\\
G^{l}_{\pm+}&=& \frac{k^{\prime2}}{L}\int^L_0dydy' \left[f^{(0)}_{L}(y,c_{l})\right]^2 G_{\pm+}(y,y')e^{-2ky'}h^2(y') \,,\nonumber\\
G^{\mu\mu}_{++} &=& \frac{k^{\prime2}}{L} \int^L_0dydy'\left[f^{(0)}_{L}(y,c_{l})\right]^2G^{}_{++}(y,y') \left[f^{(0)}_{L}(y',c_{l})\right]^2\,.
\label{ad}
\end{eqnarray}
Imposing the match conditions in Eqs. (\ref{eq:SM_observable}) and \ref{eq:RS_observable}), we obtain the relations between $g$, $g'$, $v$ and $\hat{g}$, $\hat{g}'$, $\hat{v}$ in the leading order of $v^2/k^{\prime2}$
\begin{eqnarray}
g=\hat{g}(1+d_g),\qquad
g'=\hat{g}'(1- \frac{\hat{s}^2}{\hat{c}^2}d_g),\qquad
v=\hat{v}(1+d_v)\,,
\label{couplingdeviation}
\end{eqnarray}
where $d_g$ and $d_v$ parameterize the deviations from the SM gauge couplings and Higgs VEV
\begin{eqnarray}
d_g&=&\frac{\hat{c}^2}{\hat{c}^2-\hat{s}^2}\frac{m^2_Z} {2k^{\prime2}} \left[\hat{s}^2 \left(\delta^2_{++}-\delta^2_{-+}\right) +\hat{c}^2 \left(2G^l_{++}-2G^l_{-+}-G^{\mu\mu}_{++} \right) \right] \,,
\nonumber\\
d_v&=&\frac{\hat{c}^2m^2_Z}{2k^{\prime2}} \left[ \delta^2_{++}+\delta^2_{-+} - 2G^l_{++}+2G^l_{-+}+G^{\mu\mu}_{++} \right] \,,
\label{dv_dg_expression}
\end{eqnarray}
respectively. Here $\hat{s}$ ($\hat{c}$) is the sine (cosine) function of the Weinberg angle in the SM with $\hat{s}^2=1/2- \sqrt{1-{e^2}/{(\sqrt{2}G_Fm^2_Z)}}\,\Big/2$.
We can see that the 5D Higgs VEV depends on the fermion bulk mass $c_l$. When the SM $SU(2)_L$ doublet is localized close to the UV brane, which is required to generate the charged lepton mass hierarchy, the Higgs VEV is essentially independent of $c_l$. Given the SM Higgs VEV $\hat{v}\simeq246$ GeV, we arrive at $v\simeq250$ GeV. Obviously this Higgs VEV is slightly above the SM value 246 GeV. It is generally true in the warped extra dimension due to the suppression of the $Z$ gauge boson profile near the IR brane and the additional contributions to the gauge boson masses from the wavefunction curvature terms \cite{Cacciapaglia:2006mz}. Then we turn to discussing the $Z$ boson couplings to the LH charged leptons and RH charged leptons, which are given by
\begin{eqnarray}
&&g_{L}=\frac{g}{c} \left(-\frac{1}{2}+s^2\right) \left\{1+\frac{g^2v^2}{4c^2k^{\prime2}} \left[G^{l}_{++}-
\frac{1}{\sqrt{c^2-s^2}} G^{l}_{-+}\right]\right\} \,,\nonumber\\
&&g^\alpha_{R}=\frac{gs^2}{c} \left\{1+\frac{g^2v^2}{4c^2k^{\prime2}} \left[G^{\alpha}_{++}+
\frac{\sqrt{c^2-s^2}}{s^2} G^{\alpha}_{-+}\right]\right\} \,,
\label{nc_coupling}
\end{eqnarray}
where
\begin{eqnarray}
G^{\alpha}_{\pm+}&=& \frac{k^{\prime2}}{L}\int^L_0dydy' \left[f^{(0)}_{R}(y,c_{\alpha})\right]^2 G_{\pm+}(y,y')e^{-2ky'}h^2(y') \,,
\end{eqnarray}
for $\alpha=e,\mu,\tau$, respectively. Since the RH charged leptons have different bulk masses, their couplings $g^\alpha_{R}$ are flavor-dependent. We introduce parameters $d_{L} \equiv (g^{}_{L}-\hat{g}^{}_{L})/\hat{g}^{}_{L}$ and $d^\alpha_{R} \equiv (g^{\alpha}_{R}-\hat{g}^{}_{R})/\hat{g}^{}_{R}$ to measure the deviations of the $Z$-lepton couplings from the SM values $\hat{g}^{}_{L}=\hat{g}(-1/2+\hat{s}^2)/\hat{c}$ and $\hat{g}^{}_{R}=\hat{g}\hat{s}^2/\hat{c}$.
Inserting Eq. \eqref{couplingdeviation} to Eq. \eqref{nc_coupling}, we obtain $d^{}_{L}$ and $d^\alpha_{R}$ in terms of the SM parameters and $k'$ as
\begin{eqnarray}
\nonumber d_{L} &=&  \frac{-1}{(\hat{c}^2-\hat{s}^2)^2}\frac{m^2_Z}{2k^{\prime2}} \Big[\hat{s}^2 \left(\delta^{}_{++}-\delta^{}_{-+}\right) +\hat{c}^2 \left(2G^l_{++}-2G^l_{-+}-G^{ll}_{++} \right) \Big] \\ \nonumber&&+\frac{m^2_Z}{k^{\prime2}} \Big[ G^{l}_{++}- \frac{1}{\sqrt{\hat{c}^2-\hat{s}^2}} G^{l}_{-+}\Big],\\
\nonumber d^\alpha_{R} &=& \frac{1}{(\hat{c}^2-\hat{s}^2)} \frac{m^2_Z}{2k^{\prime2}} \Big[\hat{s}^2 \left(\delta^{}_{++}-\delta^{}_{-+}\right) +\hat{c}^2 \left(2G^l_{++}-2G^l_{-+}-G^{ll}_{++} \right) \Big]\\
&&+\frac{m^2_Z}{k^{\prime2}} \Big[G^{\alpha}_{++}+ \frac{\sqrt{\hat{c}^2-\hat{s}^2}}{\hat{s}^2} G^{\alpha}_{-+}\Big] \,.\label{d_LR2}
\end{eqnarray}
The charged lepton couplings to the $Z$ boson have been measured precisely to the mille level \cite{Beringer:1900zz}. We require that both deviations $|d_{L}|$ and $|d^\alpha_{R}|$ be smaller than $2\times10^{-3}$ \cite{Csaki:2008qq}. Figure \ref{d_LR_custodial} shows the variation of $d_{L}$ and $d^\alpha_{R}$ as a function of $c_l$ and $c_\alpha$, respectively. The results of both the bulk Higgs and brane Higgs scenarios are presented. For the bulk Higgs, its mass parameter $\beta=$ 0, 0.25, 0.5 has been taken. We see that the bulk Higgs scenario is less stringently constrained than the brane Higgs, and the similar property was found in the minimal RS model \cite{Cabrer:2010si}. In most cases, the deviations are sufficiently small and decrease with the increase of $\beta$.  Both $c_l$ and $c_\alpha$ are allowed to vary in a large range. In particular, all the parameter space for fermions localized near the UV brane ($c_l>0.5$ and $c_\alpha<-0.5$) can pass the EWPM constraints.

It is remarkable that the deviation parameters $d_L$ and $d^{\alpha}_R$ are almost independent of the bulk masses of charged leptons, if they are localized close to the UV brane. These results can be understood analytically as follows. Since $c_l>0.5$ and $c_\alpha<-0.5$ are valid for UV localized charged leptons, ignoring the terms with the small factors $e^{(1-2c_l)kL}$ and $e^{(1+2c_\alpha)kL}$, we obtain
\begin{eqnarray}
&&\delta_{++}\simeq-\frac{kL(1-\beta)^2}{(2-\beta)(3-2\beta)} +\frac{(1-\beta)(3-\beta)}{2(2-\beta)^2}-\frac{1}{4kL}\,,\nonumber\\
&&\delta_{-+}\simeq-\frac{kL(1-\beta)^2}{(2-\beta)(3-2\beta)}\,, \nonumber\\
&&G^{l}_{++}\simeq G^{\alpha}_{++}\simeq \frac{(1-\beta)(3-\beta)}{4(2-\beta)^2}-\frac{1}{4kL}\,,\nonumber\\
&&
G^{\mu\mu}_{++}\simeq -\frac{1}{4kL} \,,\quad
G^{l}_{-+}\simeq G^{\alpha}_{-+}\simeq0\,.
\label{effective}
\end{eqnarray}
Although $G^{l}_{\pm+}$, $G^{\alpha}_{\pm+}$ and $G^{\mu\mu}_{++}$ are functions of fermion bulk masses generally, they become essentially independent of fermion profiles in this case \cite{Carena:2006bn}, and Eq. \eqref{d_LR2} is simplified to
\begin{eqnarray}
&&d_{L} \simeq -\frac{m^2_Z}{k^{\prime2}}\left[ \frac{(1-\beta)(3-\beta)\hat{c}^2\hat{s}^2}{(2-\beta)^2 (\hat{c}^2-\hat{s}^2)^2}+ \frac{1-8\hat{c}^2\hat{s}^2}{8kL(\hat{c}^2-\hat{s}^2)^2} \right]
\,,\nonumber\\
&&d^\alpha_{R} \simeq +\frac{m^2_Z}{k^{\prime2}} \left[\frac{(1-\beta)(3-\beta)\hat{c}^2}{2\,(2-\beta)^2(\hat{c}^2-\hat{s}^2)}- \frac{3-4\hat{s}^2}{8kL(\hat{c}^2-\hat{s}^2)} \right]  \,.
\label{d_LR_analytical}
\end{eqnarray}
Obviously $d_{L}$ and $d^\alpha_{R}$ have negative and positive contributions, respectively. For the generic scale $k'=1.5$ TeV in the custodial warped extra dimension, we obtain that both magnitudes are always smaller than the EWPM bounds $2\times10^{-3}$, as shown in Fig. \ref{d_LR_custodial}.

Furthermore, the model should also be constrained by the lepton flavor violation processes, in particularly the radiative $\mu\rightarrow e\gamma$ decay which appears at one-loop level. Due to the protection of both the $S_4$ flavor symmetry and the custodial symmetry, we expect that our model should more easily satisfy this type of constraints than the conventional warped models without flavor protection mechanisms. However, a precise and detailed calculation of the lepton flavor violations requires much work and will be presented elsewhere.

\section{\label{sec:residual}Leptonic mixing from residual symmetries within $S_4$ flavor symmetry}

In the theory of flavor symmetry, the full Lagrangian is invariant
under a family symmetry group $G_f$ at high energies above
the symmetry breaking scale. Then $G_f$ is spontaneously or explicitly broken into its subgroups $G_{\nu}$ and
$G_l$ in the neutrino and charged lepton sectors at
LO, respectively, and certain lepton mixing matrix
$U_{PMNS}$ arises as a result of the misalignment
between the two residual symmetry groups $G_{\nu}$ and $G_{l}$ \cite{Lam:2008rs}. In
general, $G_{\nu}$ and $G_l$ should be different subgroups of $G_f$,
otherwise there would be no mixing among the three mass
eigenstates. We embed the LH lepton doublets into a unitary
three-dimensional representation $\rho$ of $G_f$ to generate a
realistic lepton mixing pattern \cite{Lam:2007qc}. The invariance
under the residual symmetries $G_{\nu}$ and $G_l$ requires
\begin{equation}
\label{eq:invariance}\rho^{\dagger}(g_{\nu})m_{\nu}m^{\dagger}_{\nu}\rho(g_{\nu})=m_{\nu}m^{\dagger}_{\nu},\qquad
\rho^{\dagger}(g_{l})m_{l}m^{\dagger}_{l}\rho(g_{l})=m_{l}m^{\dagger}_{l},\quad g_{\nu}\in G_{\nu},~~g_l\in G_l\,,
\end{equation}
where the neutrino mass matrix $m_{\nu}$ and charged lepton mass
matrix $m_{l}$ are defined by
\begin{equation}
\mathcal{L}=-\overline{l_L}m_{l}l_R-\overline{\nu_L}m_{\nu}\nu_R\,,
\end{equation}
and $\l_L=(e,\mu,\tau)_L^T$, $l_R=(e,\mu,\tau)_R^T$,
$\nu_L=(\nu_e,\nu_{\mu},\nu_{\tau})_L^T$ and
$\nu_{R}=(\nu_e,\nu_{\mu},\nu_{\tau})_R^T$. Here neutrinos are assumed
to be Dirac particles. If one chooses $G_{\nu}$ or $G_l$ to be a
non-Abelian subgroup, according to the Schur's theorem, this would
result in a complete or partial degeneracy of the mass spectrum,
which is not compatible with the observation that both the light neutrinos and the charged leptons have different masses. Therefore
we shall restrict $G_{\nu}$ and $G_l$ to be Abelian subgroups. The representation matrices $\rho(g_{\nu})$ and $\rho(g_{e})$ can be diagonalized by two unitary transformations $U_{\nu}$ and $U_{e}$ as follows
\begin{equation}
U^{\dagger}_{\nu}\rho(g_{\nu})U_{\nu}=\widehat{\rho}(g_{\nu}),\qquad  U^{\dagger}_{l}\rho(g_{l})U_{l}=\widehat{\rho}(g_{l})\,,
\end{equation}
where $\widehat{\rho}(g_{\nu})$ and $\widehat{\rho}(g_{l})$ are diagonal unitary matrices, and $g_{\nu}$ and $g_l$ can be taken to be the generators of the Abelian subgroups of $G_{\nu}$ and $G_l$, respectively. We note that the matrices $U_{\nu}$ and $U_{l}$ are determined uniquely up to diagonal unitary matrices $K_{\nu,l}$ and permutation matrices $P_{\nu,l}$, respectively
\begin{equation}
U_{\nu,l}\rightarrow U_{\nu,l}K_{\nu,l}P_{\nu,l}\,.
\end{equation}
From the invariance requirement of Eq. (\ref{eq:invariance}), it follows
\begin{eqnarray}
\nonumber&U^{\dagger}_{\nu}m_{\nu}m^{\dagger}_{\nu}U_{\nu}=U^{\dagger}_{\nu}\rho^{\dagger}(g_{\nu})m_{\nu}m^{\dagger}_{\nu}\rho(g_{\nu})U_{\nu}
=\widehat{\rho}^{\;\dagger}(g_{\nu})U^{\dagger}_{\nu}m_{\nu}m^{\dagger}_{\nu}U_{\nu}\widehat{\rho}(g_{\nu})\,,&\\
&U^{\dagger}_{l}m_{l}m^{\dagger}_{l}U_{l}=U^{\dagger}_{l}\rho^{\dagger}(g_{l})m_{l}m^{\dagger}_{l}\rho(g_{l})U_{l}
=\widehat{\rho}^{\;\dagger}(g_{l})U^{\dagger}_{l}m_{l}m^{\dagger}_{l}U_{l}\widehat{\rho}(g_{l})\,.&
\end{eqnarray}
As a result, both $U^{\dagger}_{\nu}m_{\nu}m^{\dagger}_{\nu}U_{\nu}$ and $U^{\dagger}_{l}m_{l}m^{\dagger}_{l}U_{l}$ have to be diagonal, i.e.,
the hermitian combination $m_{\nu}m^{\dagger}_{\nu}$ is diagonalized by $U_{\nu}$, and $m_{l}m^{\dagger}_{l}$ is diagonalized by $U_{l}$ as well. Consequently, we can determine the lepton mixing matrix $U_{PMNS}$ as
\begin{equation}
U_{PMNS}=U^{\dagger}_{l}U_{\nu}
\end{equation}
up to permutations of rows and columns. Obviously if $G_{l}$ and $G_{\nu}$ are the same subgroups of $S_4$, the PMNS matrix would be an identity matrix, which is incompatible with the present data \cite{Tortola:2012te,Fogli:2012ua,GonzalezGarcia:2012sz}. In the following, we perform a comprehensive study of the possible lepton mixing patterns within $S_4$ flavor symmetry by scanning the different choices for $G_{\nu}$ and $G_l$. The non-trivial Abelian subgroups of $S_4$ are $Z_2$, $Z_3$, $Z_4$ and $K_4$, where $K_4\cong Z_2\times Z_2$ is the Klein four group. The $Z_2$ subgroups are irrelevant to us, since at least two of the eigenvalues are degenerate, such that the three generations of neutrinos or charged leptons can not be distinguished. On the technique side, this degeneracy prevents us from uniquely pinning down the lepton mixing matrix, because the unitary transformations $U_{\nu}$ and $U_l$ can not be fixed unambiguously in this case. The group structure of $S_4$ has been discussed in Ref. \cite{Ding:2009iy} in detail, and concretely it has four $Z_3$ subgroups, three $Z_4$ subgroups and four $K_4$ subgroups. In terms of the generators $S$ and $T$, they can be expressed as
\begin{itemize}
\item{$Z_3$ subgroups}
\begin{equation}
\begin{array}{ll}
Z^{(1)}_3=\left\{1,T,T^2\right\},  & \qquad Z^{(2)}_3=\left\{1,T^2S^2,S^2T\right\}, \\
Z^{(3)}_3=\left\{1,S^2TS^2,STS\right\},& \qquad Z^{(4)}_3=\left\{1,S^2T^2,TS^2\right\}\,.
\end{array}
\end{equation}
\item{$Z_4$ subgroups}
\begin{eqnarray*}
Z^{(1)}_4=\left\{1,S,S^2,S^3\right\},\quad Z^{(2)}_4=\left\{1,TS,T^2S^2T,T^2ST\right\},\quad Z^{(3)}_4=\left\{1,ST,TS^2T^2,TST^2\right\}\,.
\end{eqnarray*}
\item{$K_4$ subgroups}
\begin{equation}
\begin{array}{ll}
K^{(1)}_4=\left\{1,STS^2,T^2S,TS^2T^2\right\},  & \qquad K^{(2)}_4=\left\{1,TSTS^2,TST,S^2\right\}, \\
K^{(3)}_4=\left\{1,ST^2,S^2TS,T^2S^2T\right\}, & \qquad K^{(4)}_4=\left\{1,TS^2T^2,S^2,T^2S^2T\right\}\,.
\end{array}
\end{equation}
\end{itemize}
In the following, all the possible Abelian remnant symmetry groups $G_{\nu}$ and $G_l$ of $S_4$ are considered one by one, the corresponding lepton mixing matrix $U_{PMNS}$ and the group structure generated by $G_{\nu}$ and $G_l$ are listed in Table \ref{tab:residual_symmetry}. Since we are mainly interested in the mixing angles, we only present the absolute values of the mixing matrix entries, which are denoted by $||U_{PMNS}||$. From this table, we see that we have a broader set of remnant symmetry groups $G_{l}$ and $G_{\nu}$ for Dirac neutrinos, recalling that $G_{\nu}$ is generally fixed to $K_4$ in the case of Majorana neutrinos. Clearly seven leptonic mixing patterns up to permutations of rows and columns can be produced within $S_4$, and only four of them remain if we require the elements of $G_{l}$ and $G_{\nu}$ generate the whole group $S_4$. It is interesting to note that if the $S_4$ flavor symmetry is broken into $G_{\nu}=Z_4$, $G_{l}=Z_3$ or $G_{\nu}=Z_3$, $G_{l}=Z_4$, where $Z_3$ and $Z_4$ denote any $Z_3$ and $Z_4$ subgroups, respectively, we will be able to produce the so-called TFH mixing pattern \cite{Toorop:2011jn,Ding:2012xx,King:2012in}. The same point has also been observed in Ref.~\cite{Toorop:2011jn} ~\footnote{Only the choice $G_{\nu}=Z_4$ and $G_{\l}=Z_3$ was pointed out in Ref.~\cite{Toorop:2011jn}.}. However, if neutrinos are Majorana particles, to generate TFH mixing without fine tuning, the flavor symmetry group should be $\Delta(96)$ or any group containing it. This implies that, in contrast to $\Delta(600)$, $\left(Z_{18}\times Z_6\right)\rtimes S_3$ and $\Delta(1536)$ for Majorana neutrinos, smaller family symmetry groups could possibly yield lepton mixing patterns which lie within the $3\sigma$ ranges of the current global fitting values, if neutrinos are Dirac particles~\cite{progress}. Similar to the Majorana neutrinos case, the combinations of $G_{\nu}=K^{(1,2,3)}_4$ and $G_{l}=Z_3$ lead to the tri-bimaximal mixing. On the other hand, we are able to derive the DC mixing for $G_{\nu}=Z_3$ and $G_{l}=K^{(1,2,3)}_4$, which is exactly the LO symmetry breaking chain of our model in section \ref{sec:model}. Finally, the bimaximal mixing can be produced if the residual symmetry groups in the neutrino and charged lepton sectors are $K_4$ or $Z_4$, please refer to Table \ref{tab:residual_symmetry} for detailed possible choices of $G_{\nu}$ and $G_l$.

\begin{center}
\begin{table}[t!]
\resizebox{\textwidth}{!}{
\begin{tabular}{|c|c|c|c|c|}  \hline\hline
 & \multirow{2}{*}{${\tt ||U_{PMNS}|| } $ } & \multirow{2}{*}{\tt Mixing Angles} & {\tt Symmetry Breaking}  & {\tt Generated }\\

 &   &  & $\Big\langle G_{\nu},G_l\Big\rangle,G_{\nu}\neq G_l$ & {\tt Group} \\  \hline

  &   &    &     &      \\ [-0.15in]

I & $\frac{1}{3}\left(\begin{array}{ccc}
2 & 2 & 1 \\
2 & 1 & 2  \\
1 & 2 & 2
\end{array}\right)$ & $\begin{array}{c}
\sin^2\theta_{13}=\frac{1}{9} \\
\sin^2\theta_{12}=\frac{1}{2} \\
\sin^2\theta_{23}=\frac{1}{2}
\end{array}$  &  $\Big\langle Z_3, Z_3\Big\rangle $  &  \multirow{4}{*}{$A_4$} \\ [0.25in] \cline{1-4}

  &   &    &     &      \\ [-0.15in]

II  & $\frac{1}{\sqrt{3}}\left(\begin{array}{ccc}
1 &  1  &  1  \\
1 &  1  &  1  \\
1 &  1  &  1
\end{array}\right)$   &  $\begin{array}{c}
\sin^2\theta_{13}=\frac{1}{3}  \\
\sin^2\theta_{12}=\frac{1}{2}  \\
\sin^2\theta_{23}=\frac{1}{2}
\end{array}$  &   $\Big\langle Z_3, K^{(4)}_4\Big\rangle, \Big\langle K^{(4)}_4,Z_3 \Big\rangle$  &   \\  [0.25in] \hline

  &   &    &     &      \\ [-0.15in]

III   &  $\frac{1}{6}\left(\begin{array}{ccc}
3+\sqrt{3}  &   2\sqrt{3}  &  3-\sqrt{3} \\
3-\sqrt{3}   &  2\sqrt{3}    & 3+\sqrt{3}  \\
2\sqrt{3}     &   2\sqrt{3}  & 2\sqrt{3}
\end{array}\right) $ &   $\begin{array}{c}
\sin^2\theta_{13}=\frac{1}{6}(2-\sqrt{3}) \\
\sin^2\theta_{12}=\frac{1}{13}(8-2\sqrt{3}) \\
\sin^2\theta_{23}=\frac{1}{13}(5+2\sqrt{3})
\end{array}$ & $\Big\langle Z_3, Z_4 \Big\rangle, \Big\langle Z_4,Z_3 \Big\rangle$   &  \multirow{12}{*}{$S_4$}   \\  [0.25in] \cline{1-4}

  &   &    &     &      \\ [-0.15in]

IV  & $\frac{1}{\sqrt{6}}\left(\begin{array}{ccc}
2  &  \sqrt{2}   &  0  \\
1  &  \sqrt{2}   &  \sqrt{3}  \\
1  &  \sqrt{2}   &  \sqrt{3}
\end{array}\right)$  &  $\begin{array}{c}
\sin^2\theta_{13}=0  \\
\sin^2\theta_{12}=\frac{1}{3}  \\
\sin^2\theta_{23}=\frac{1}{2}
\end{array}$   &  $\Big\langle K^{(1,2,3)}_4,Z_3\Big\rangle$  &  \\   [0.25in] \cline{1-4}

  &   &    &     &      \\ [-0.15in]

V  & $\frac{1}{\sqrt{6}}\left(\begin{array}{ccc}
\sqrt{3}  &  \sqrt{3}  &  0  \\
1     &  1   &  2  \\
\sqrt{2} &  \sqrt{2}   &  \sqrt{2}
\end{array}\right)$ &  $\begin{array}{c}
\sin^2\theta_{13}=0   \\
\sin^2\theta_{12}=\frac{1}{2}   \\
\sin^2\theta_{23}=\frac{2}{3}
\end{array}$   & $\Big\langle Z_3,K^{(1,2,3)}_4 \Big\rangle$ &   \\  [0.25in] \cline{1-4}

  &   &    &     &      \\ [-0.15in]

VI  & $\frac{1}{2}\left(\begin{array}{ccc}
\sqrt{2}   &   \sqrt{2}   &  0\\
1          &     1        & \sqrt{2}  \\
1          &     1        & \sqrt{2}
\end{array}\right)$  & $\begin{array}{c}
\sin^2\theta_{13}=0\\
\sin^2\theta_{12}=\frac{1}{2}  \\
\sin^2\theta_{23}=\frac{1}{2}
\end{array}$    &  $\begin{array}{cc}
\Big\langle K^{(1,3)}_4,Z^{(1)}_4\Big\rangle, & \Big\langle K^{(1,2)}_4,Z^{(2)}_4\Big\rangle,  \\
     &      \\ [-0.18in]
\Big\langle K^{(2,3)}_4, Z^{(3)}_4\Big\rangle, &  \Big\langle Z^{(1)}_4,K^{(1,3)}_4 \Big\rangle,  \\
     &      \\ [-0.18in]
\Big\langle Z^{(2)}_4, K^{(1,2)}_4\Big\rangle,  &  \Big\langle Z^{(3)}_4,K^{(2,3)}_4 \Big\rangle , \\
     &      \\ [-0.18in]
\Big\langle Z_4,Z_4 \Big\rangle, &  \Big\langle K^{(1,2,3)}_4,K^{(1,2,3)}_4 \Big\rangle  \\
\end{array}$   &    \\ [0.60in] \hline

  &   &    &     &      \\ [-0.15in]

VII  &  $\frac{1}{\sqrt{2}}\left(\begin{array}{ccc}
\sqrt{2}   &   0   &   0 \\
0          &   1   &   1 \\
0          &   1   &   1
\end{array}\right)$ & $\begin{array}{c}
\sin^2\theta_{13}=0\\
\sin^2\theta_{12}=0  \\
\sin^2\theta_{23}=\frac{1}{2}
\end{array}$    &   $\begin{array}{cc}
\Big\langle K^{(2,4)}_4, Z^{(1)}_4\Big\rangle, &  \Big\langle K^{(3,4)}_4, Z^{(2)}_4\Big\rangle, \\
     &      \\ [-0.18in]
\Big\langle K^{(1,4)}_4, Z^{(3)}_4 \Big\rangle, &  \Big\langle Z^{(1)}_4,K^{(2,4)}_4 \Big\rangle, \\
     &      \\ [-0.18in]
\Big\langle Z^{(2)}_4,K^{(3,4)}_4\Big\rangle, &  \Big\langle Z^{(3)}_4,K^{(1,4)}_4 \Big\rangle, \\
     &      \\ [-0.18in]
\Big\langle K^{(4)}_4,K^{(1,2,3)}_4 \Big\rangle, &  \Big\langle K^{(1,2,3)}_4,K^{(4)}_4 \Big\rangle
\end{array}$    &  $D_4$  \\  [0.60in] \hline\hline

\end{tabular}
}
\caption{\label{tab:residual_symmetry}The leptonic mixing patterns and the generated group structures for the $S_4$ flavor symmetry breaking into different subgroups $G_{\nu}$ and $G_l$ in the neutrino and charged lepton sectors, respectively, where $Z_3$ ($Z_4$) denotes any $Z_3$ ($Z_4$) subgroup. $D_4$ stands for the dihedral group of degree 4 and order 8, and geometrically it is the symmetry group of the square.}
\end{table}
\end{center}

\section{\label{sec:conclusion} Summary and conclusions }

The remarkable discovery of a rather large mixing angle $\theta_{13}$ and the recent indication of a sizable deviation of the atmospheric neutrino
mixing angle $\theta_{23}$ from $\pi/4$ have profound impact on neutrino physics. Many neutrino mass models are ruled out due to their predictions of small $\theta_{13}$. In the present work, from the viewpoint of flavor symmetry, we have investigated the proposal that the LO neutrino mixing matrix is the DC pattern and then it is corrected by the NLO contributions of order $\lambda_c$ such that the resulting lepton mixing angles can be in the experimentally preferred ranges.

It is well-known that warped extra dimensions provide an attractive solution to the gauge hierarchy problem, an understanding of the observed hierarchies in the fermion masses and mixing angles, and a natural suppression of the dangerous flavor changing neutral current processes. Inspired by the successes of discrete flavor symmetry in predicting the lepton mixing angles, we combine the warped extra dimension and the discrete flavor symmetry together in this work. We construct a lepton model based on the flavor symmetry $S_4\times Z_2\times Z'_2$ in the  warped extra dimension with the custodial symmetry $SU(2)_L\times SU(2)_R\times U(1)_X\times P_{LR}$. In this model, all the matter fields, gauge fields and the Higgs field are allowed to propagate in the bulk. The flavon fields breaking the $S_4$ flavor symmetry in the neutrino and charged lepton sectors live on the UV and IR branes, respectively. The light neutrinos are taken to be Dirac particles, the correct neutrino mass scale can be obtained by choosing the fermion and Higgs bulk mass parameters, and the charged lepton mass hierarchy is generated as usual in RS models via wavefunction overlaps. The lepton mixing matrix turns out to be of the DC form at LO, and the neutrino mass spectrum is predicted to be inverted hierarchy.

For DC mixing under the hypothesis of Dirac neutrinos, although the symmetry group of the left-handed neutrino fields is $\mathcal{G}_{\nu}\equiv U(1)\times U(1)\times U(1)\times U(1)$ in the charged lepton diagonal basis, it is sufficient to produce DC mixing by just requiring that the neutrino sector be invariant under the action of one element $h\in\mathcal{G}_{\nu}$ with non-degenerate eigenvalues~\footnote{This statement is generally applicable to the mass-independent textures for Dirac neutrinos.}. Therefore the DC mixing can be generated if the flavor symmetry group $G_f$ contains the element $h$ which can be preserved in the neutrino sector. This is the reason why DC mixing can be derived from the $S_4$ flavor symmetry, and $h$ corresponds to the generator $T$ in the present work, as shown in section \ref{sec:DC}.

Although the exact DC mixing leads to the mixing angles $\sin^2\theta^{DC}_{13}=0$, $\sin^2\theta^{DC}_{12}=1/2$ and $\sin^2\theta^{DC}_{23}=2/3$, which is excluded by the present data, the LO predictions are modified by the high dimensional operators allowed by the symmetries of the model. The induced deviations of all the three mixing angles from the DC values are linear in the expansion parameter $V/\Lambda$, where $V$ denotes the VEV of any flavon field. It is usually assumed that all the VEVs are approximately of the same order of magnitude. In order to achieve agreement with the experimental measurements, the parameter $V/\Lambda$ should be of order $\lambda_c$. Detailed numerical simulations show that the neutrino mass spectrum is still inverted hierarchy and the second octant of $\theta_{23}$ is preferred after the NLO contributions are taken into account. Therefore our model can be tested in next generation neutrino oscillation experiments, which are deliberately designed to resolve the neutrino mass ordering, the octant of $\theta_{23}$ and eventually determine the leptonic CP violating phases. Since the first octant of $\theta_{23}$ is also allowed by the present data, we suggest that the ``modified" DC (MDC) mixing as a LO approximation and subsequent $\mathcal{O}(\lambda_c)$ corrections can account for the data on neutrino mixing for the case of $\theta_{23}$ in the first octant, where the MDC mixing matrix can be obtained by permuting the second and the third rows of the DC mixing matrix. By exchanging the assignments for $\xi^{2}_3$ and $\xi^{3}_3$ and unifying $\left(\xi^{1}_1,\xi^{3}_1,\xi^{2}_1\right)$ instead of $\left(\xi^{1}_1,\xi^{2}_1,\xi^{3}_1\right)$ into a $S_4$ triplet $\mathbf{3}$, our constructed DC model would become a model for the MDC mixing. Moreover, we study the constraints on the model from the electroweak precision measurements, which are estimated by requiring small deviations from the SM tree-level couplings. We show that the model is compatible with the experimental constraints for usual KK mass scales of order 3 TeV.

Finally we have investigated the associated lepton mixing patterns for all the possible remnant subgroups $G_{\nu}$ and $G_{l}$ in the neutrino and charged lepton sectors respectively within $S_4$, where neutrinos are assumed to be Dirac particles. It is notable that the TFH mixing can be generated within $S_4$ instead of $\Delta(96)$ for Dirac neutrinos if the $S_4$ flavor symmetry is broken into $G_{\nu}=Z_4$, $G_{l}=Z_3$ or $G_{\nu}=Z_3$, $G_{l}=Z_4$. 
We can derive the DC mixing from the combinations of $G_{\nu}=Z_3$ and $G_{l}=K^{(1,2,3)}_4$. This is the guiding principle of our model building. In addition, the tri-bimaximal mixing and bimaximal mixing can also be produced if we require the elements of $G_{\nu}$ and $G_{l}$ give rise to the entire group $S_4$.

\section*{Acknowledgements}
This work is initiated from the simulating discussion with Prof. Zhi-zhong Xing, and we are very grateful for his continuous support, encouragement and correcting the manuscript. The research was partially supported by the National Natural Science Foundation of China under Grant Nos. 11275188, 11179007 and 11135009.

\newpage


\appendix

\section*{\label{apd:a} \qquad\qquad Appendix A: The group theory of $S_4$}

$S_4$ is a symmetric group of degree four. Geometrically, it is the
group of orientation-preserving symmetries of the cube (or
equivalently, the octahedron). The group has 24 distinct elements,
and it can be generated by two generators $S$ and $T$ which satisfy
the relations
\begin{equation}
\label{4} S^4=T^3=1,~~~ST^2S=T\,.
\end{equation}
Without loss of generality, we could choose
\begin{equation}
\label{5}S=(1234),~~~~~T=(123)
\end{equation}
where the cycle (1234) denotes the permutation
$(1,2,3,4)\rightarrow(2,3,4,1)$, and (123) means
$(1,2,3,4)\rightarrow(2,3,1,4)$. The 24 group elements are divided
into 5 conjugacy classes as follows:
\begin{eqnarray*}
{\cal C}_1:\;&& 1\\
{\cal C}_2:\;&& STS^2=(12),\;TSTS^2=(13),\;ST^2=(14),\;S^2TS=(23),\;TST=(24),\;T^2S=(34)\\
{\cal C}_3:\;&&TS^2T^2=(12)(34),\;S^2=(13)(24),\;T^2S^2T=(14)(23)\\
{\cal C}_4:\;&& T=(123),\;T^2=(132),\;T^2S^2=(124),\;S^2T=(142),\;S^2TS^2=(134),\;STS=(143),\\
&&S^2T^2=(234),\;TS^2=(243)\\
{\cal C}_5:\;&& S=(1234),\;T^2ST=(1243),\;ST=(1324),\;\;TS=(1342),\;TST^2=(1423),\;\\
&&S^3=(1432)\\
\end{eqnarray*}
\begin{table}[t]
\begin{center}
\begin{tabular}{|c|c|c|c|c|c|c|c|}\hline\hline
              &   $~h_{{\cal C}_i}~$   &   $~n_{{\cal C}_i}~$    &    $~\chi_{\mathbf{1}}~$    &   $~\chi_{\mathbf{1'}}~$    &     $~\chi_{\mathbf{2}}~$    &    $~\chi_{\mathbf{3}}~$    &    $~\chi_{\mathbf{3}'}~$   \\ \hline
$~\mathcal{C}_1~$    &    1        &      1               &     1                     &       1                   &      2                     &        3                  &    3  \\

$~\mathcal{C}_2~$    &    2        &      6               &     1                     &       $-1$                &      0                     &        1                  &    $-1$  \\

$~\mathcal{C}_3~$    &    2        &      3               &     1                     &       1                   &      2                     &        $-1$               &    $-1$  \\

$~\mathcal{C}_4~$    &    3        &      8               &     1                     &       1                   &      $-1$                  &         0                 &    0   \\

$~\mathcal{C}_5~$    &    4        &      6               &     1                     &      $-1$                 &      0                     &        $-1$               &   1  \\\hline\hline
\end{tabular}
\caption{\label{tab:character} Character table of $S_4$ group, where $h_{{\cal C}_i}$ denotes the order of the elements of ${\cal C}_i$, and $n_{{\cal C}_i}$ is the number of the elements in the class ${\cal C}_i$.}
\end{center}
\end{table}
Since the number of the irreducible representations is equal to the number of conjugacy classes, $S_4$ has five irreducible representations: two singlets $\mathbf{1}$ and $\mathbf{1'}$, one doublet $\mathbf{2}$, and two triplets $\mathbf{3}$ and $\mathbf{3}'$.
We note that both $\mathbf{3}$ and $\mathbf{3}'$ are faithful representations. The character table of $S_4$ is presented in Table \ref{tab:character}, and the Kronecker products between two irreducible representations, which can be easily deduced from the character table, are listed below:
\begin{eqnarray}
\nonumber&&\mathbf{1}\otimes R=R\otimes
\mathbf{1}=R,~~\mathbf{1'}\otimes\mathbf{1'}=\mathbf{1},~~\mathbf{1'}\otimes\mathbf{2}=\mathbf{2},~~\mathbf{1'}\otimes\mathbf{3}=\mathbf{3}',~~\mathbf{1'}\otimes\mathbf{3}'=\mathbf{3},\\
\nonumber&&\mathbf{2}\otimes\mathbf{2}=\mathbf{1}\oplus\mathbf{1'}\oplus\mathbf{2},~~\mathbf{2}\otimes\mathbf{3}=\mathbf{3}\oplus\mathbf{3}',~~\mathbf{2}\otimes\mathbf{3}'=\mathbf{3}\oplus\mathbf{3}',~~
\mathbf{3}\otimes\mathbf{3}=\mathbf{1}\oplus\mathbf{2}\oplus\mathbf{3}\oplus\mathbf{3}',\\
&&\mathbf{3}\otimes\mathbf{3}'=\mathbf{1'}\oplus\mathbf{2}\oplus\mathbf{3}\oplus\mathbf{3}',~~\mathbf{3}'\otimes\mathbf{3}'=\mathbf{1}\oplus\mathbf{2}\oplus\mathbf{3}\oplus\mathbf{3}'\,,
\end{eqnarray}
where $R$ denotes any $S_4$ irreducible representation. Different presentations of the $S_4$ group have been discussed in literatures \cite{Ding:2009iy,Ma:2005pd,Hagedorn:2006ug,Bazzocchi:2009pv,Hagedorn:2010th}. In the present paper, we would like to work in the charged lepton diagonal basis such that the DC mixing completely comes from the neutrino sector, the representation matrices of the generators $S$ and $T$ for different $S_4$ irreducible representations are presented in Table \ref{tab:representation_matrix}. The adopted representation here is related to other existing choices by unitary transformations. In the following, we list the Clebsch-Gordan coefficients derived from this basis. All the Clebsch-Gordan coefficients are reported in the form of $\alpha\otimes\beta$, where the $\alpha_i$ are the base vectors of the representation on the left of the product, and the $\beta_j$ are from the representation on the right of the product.

\begin{table}[t!]
\begin{center}
\begin{tabular}{|c|c|c|} \hline\hline
  &  $S$   &  $T$    \\ \hline

$~\mathbf{1}~$  & 1  & 1  \\  \hline

$~\mathbf{1'}~$  & $-1$  & 1  \\  \hline

  &    &       \\ [-0.20in]

$~\mathbf{2}~$ & $\left(
\begin{array}{cc}
 1 & 0 \\
 0 & -1
\end{array}
\right)$  & $\frac{1}{2}\left(
\begin{array}{cc}
 -1 & \sqrt{3} \\
 -\sqrt{3} & -1
\end{array}
\right)$   \\ [0.12in] \hline

  &    &       \\ [-0.20in]

$~\mathbf{3}~$  & $\left(
\begin{array}{ccc}
 0 & -1 & 0 \\
 1 & 0 & 0 \\
 0 & 0 & -1
\end{array}
\right)$  &   $\frac{1}{2}\left(
\begin{array}{ccc}
 -1 & -1 & \sqrt{2} \\
 1 & 1 & \sqrt{2} \\
 -\sqrt{2} & \sqrt{2} & 0
\end{array}
\right)$   \\  [0.12in] \hline

  &    &       \\ [-0.20in]

$\mathbf{3}'$  & $\left(
\begin{array}{ccc}
 0 & 1 & 0 \\
 -1 & 0 & 0 \\
 0 & 0 & 1
\end{array}
\right)$  &   $\frac{1}{2}\left(
\begin{array}{ccc}
 -1 & -1 & \sqrt{2} \\
 1 & 1 & \sqrt{2} \\
 -\sqrt{2} & \sqrt{2} & 0
\end{array}
\right)$   \\ [0.12in] \hline\hline

\end{tabular}
\caption{\label{tab:representation_matrix} Representation
matrices for the generators $S$ and $T$ in different $S_4$
irreducible representations.}
\end{center}
\end{table}

\begin{itemize}

\item{$\mathbf{1'}\otimes\mathbf{2}=\mathbf{2}$}

\begin{equation}
\nonumber\mathbf{2}\sim\left(
\begin{array}{c}
 \alpha _1 \beta _2 \\
 -\alpha _1 \beta _1
\end{array}
\right)
\end{equation}


\item{$\mathbf{1'}\otimes\mathbf{3}=\mathbf{3}'$}

\begin{equation}
\nonumber\mathbf{3}'\sim\left(
\begin{array}{c}
 \alpha _1 \beta _1 \\
 \alpha _1 \beta _2 \\
 \alpha _1 \beta _3
\end{array}
\right)
\end{equation}


\item{$\mathbf{1'}\otimes\mathbf{3}'=\mathbf{3}$}

\begin{equation}
\nonumber\mathbf{3}\sim\left(
\begin{array}{c}
 \alpha _1 \beta _1 \\
 \alpha _1 \beta _2 \\
 \alpha _1 \beta _3
\end{array}
\right)
\end{equation}


\item{$\mathbf{2}\otimes\mathbf{2}=\mathbf{1}\oplus\mathbf{1'}\oplus\mathbf{2}$}

\begin{equation}
\nonumber\mathbf{1}\sim
 \alpha _1 \beta _1+\alpha _2 \beta _2
\end{equation}

\begin{equation}
\nonumber\mathbf{1'}\sim
 \alpha _1 \beta _2-\alpha _2 \beta _1
\end{equation}

\begin{equation}
\nonumber\mathbf{2}\sim\left(
\begin{array}{c}
 \alpha _2 \beta _2-\alpha _1 \beta _1 \\
 \alpha _1 \beta _2+\alpha _2 \beta _1
\end{array}
\right)
\end{equation}


\item{$\mathbf{2}\otimes\mathbf{3}=\mathbf{3}\oplus\mathbf{3}'$}

\begin{equation}
\nonumber\mathbf{3}\sim\left(
\begin{array}{c}
 \alpha _1 \beta _1+\sqrt{3} \alpha _2 \beta _2 \\
 \alpha _1 \beta _2+\sqrt{3} \alpha _2 \beta _1 \\
 -2 \alpha _1 \beta _3
\end{array}
\right)
\end{equation}

\begin{equation}
\nonumber\mathbf{3}'\sim\left(
\begin{array}{c}
 \sqrt{3}\alpha _1 \beta _2-\alpha _2 \beta _1 \\
 \sqrt{3}\alpha _1 \beta _1-\alpha _2 \beta _2 \\
  2 \alpha _2 \beta _3
\end{array}
\right)
\end{equation}


\item{$\mathbf{2}\otimes\mathbf{3}'=\mathbf{3}\oplus\mathbf{3}'$}

\begin{equation}
\nonumber\mathbf{3}\sim\left(
\begin{array}{c}
 \sqrt{3}\alpha _1 \beta _2-\alpha _2 \beta _1 \\
 \sqrt{3}\alpha _1 \beta _1-\alpha _2 \beta _2 \\
  2 \alpha _2 \beta _3
\end{array}
\right)
\end{equation}

\begin{equation}
\nonumber\mathbf{3}'\sim\left(
\begin{array}{c}
 \alpha _1 \beta _1+\sqrt{3} \alpha _2 \beta _2 \\
 \alpha _1 \beta _2+\sqrt{3} \alpha _2 \beta _1 \\
 -2 \alpha _1 \beta _3
\end{array}
\right)
\end{equation}


\item{$\mathbf{3}\otimes\mathbf{3}=\mathbf{1}\oplus\mathbf{2}\oplus\mathbf{3}\oplus\mathbf{3}'$}

\begin{equation}
\nonumber\mathbf{1}\sim\alpha _1 \beta _1+\alpha _2 \beta _2+\alpha _3 \beta _3
\end{equation}

\begin{equation}
\nonumber\mathbf{2}\sim\left(
\begin{array}{c}
 \alpha _1 \beta _1+\alpha _2 \beta _2-2 \alpha _3 \beta _3 \\
 \sqrt{3} \alpha _1 \beta _2+\sqrt{3} \alpha _2 \beta _1
\end{array}
\right)
\end{equation}

\begin{equation}
\nonumber\mathbf{3}\sim\left(
\begin{array}{c}
 \alpha _1 \beta _3+\alpha _3 \beta _1 \\
  -\alpha _2 \beta _3-\alpha _3 \beta _2 \\
 \alpha _1 \beta _1-\alpha _2 \beta _2
\end{array}
\right)
\end{equation}

\begin{equation}
\nonumber\mathbf{3}'\sim\left(
\begin{array}{c}
 \alpha _2 \beta _3-\alpha _3 \beta _2 \\
 \alpha _3 \beta _1-\alpha _1 \beta _3 \\
 \alpha _1 \beta _2-\alpha _2 \beta _1
\end{array}
\right)
\end{equation}


\item{$\mathbf{3}\otimes\mathbf{3}'=\mathbf{1'}\oplus\mathbf{2}\oplus\mathbf{3}\oplus\mathbf{3}'$}

\begin{equation}
\nonumber\mathbf{1'}\sim \alpha _1 \beta _1+\alpha _2 \beta _2+\alpha _3 \beta _3
\end{equation}

\begin{equation}
\nonumber\mathbf{2}\sim\left(
\begin{array}{c}
  -\sqrt{3} \alpha _1 \beta _2-\sqrt{3} \alpha _2 \beta _1 \\
 \alpha _1 \beta _1+\alpha _2 \beta _2-2 \alpha _3 \beta _3
\end{array}
\right)
\end{equation}

\begin{equation}
\nonumber\mathbf{3}\sim\left(
\begin{array}{c}
 \alpha _2 \beta _3-\alpha _3 \beta _2 \\
 \alpha _3 \beta _1-\alpha _1 \beta _3 \\
 \alpha _1 \beta _2-\alpha _2 \beta _1
\end{array}
\right)
\end{equation}

\begin{equation}
\nonumber\mathbf{3}'\sim\left(
\begin{array}{c}
 \alpha _1 \beta _3+\alpha _3 \beta _1 \\
  -\alpha _2 \beta _3-\alpha _3 \beta _2 \\
 \alpha _1 \beta _1-\alpha _2 \beta _2
\end{array}
\right)
\end{equation}


\item{$\mathbf{3}'\otimes\mathbf{3}'=\mathbf{1}\oplus\mathbf{2}\oplus\mathbf{3}\oplus\mathbf{3}'$}

\begin{equation}
\nonumber\mathbf{1}\sim\alpha _1 \beta _1+\alpha _2 \beta _2+\alpha _3 \beta _3
\end{equation}

\begin{equation}
\nonumber\mathbf{2}\sim\left(
\begin{array}{c}
 \alpha _1 \beta _1+\alpha _2 \beta _2-2 \alpha _3 \beta _3 \\
 \sqrt{3} \alpha _1 \beta _2+\sqrt{3} \alpha _2 \beta _1
\end{array}
\right)
\end{equation}

\begin{equation}
\nonumber\mathbf{3}\sim\left(
\begin{array}{c}
 \alpha _1 \beta _3+\alpha _3 \beta _1 \\
 -\alpha _2 \beta _3-\alpha _3 \beta _2 \\
 \alpha _1 \beta _1-\alpha _2 \beta _2
\end{array}
\right)
\end{equation}

\begin{equation}
\nonumber\mathbf{3}'\sim\left(
\begin{array}{c}
 \alpha _2 \beta _3-\alpha _3 \beta _2 \\
 \alpha _3 \beta _1-\alpha _1 \beta _3 \\
 \alpha _1 \beta _2-\alpha _2 \beta _1
\end{array}
\right)
\end{equation}

\end{itemize}


\section*{\label{apd:b}  Appendix B: The flavon potential and vacuum alignment }
In the context of extra dimensions, the vacuum alignment problem is greatly simplified~\cite{Altarelli:2005yp}, since each flavon field is usually assumed to live on a 4D brane either at $y=0$ or at $y=L$, as listed in Table~\ref{tab:field}. The most general renormalizable flavon potential consistent with the flavor symmetry $S_4\times Z_2\times Z'_2 $ is given by
\begin{equation} 
\mathcal{V}=V_{y=0}+V_{y=L}
\end{equation}
where
\begin{eqnarray}
\nonumber&&V_{y=0}=V(\zeta)+V(\phi)+V(\rho)+V(\zeta,\phi)+V(\zeta,\rho)+V(\phi,\rho)\,,\\
&&V_{y=L}=V(\chi)+V(\varphi)+V(\chi,\varphi)\,,
\end{eqnarray}
with
\begin{eqnarray}
\nonumber&&V(\zeta)=M^2_{\zeta}\zeta^2+f_1\zeta^4\,,\\
\nonumber&&V(\phi)=M^2_{\phi}\left(\phi\phi\right)_{\mathbf{1}}+f_2\left(\phi\phi\right)_{\mathbf{1}}\left(\phi\phi\right)_{\mathbf{1}}+
f_3\left(\left(\phi\phi\right)_{\mathbf{2}}\left(\phi\phi\right)_{\mathbf{2}}\right)_{\mathbf{1}}\\
\nonumber&&\qquad\quad+f_4\left(\left(\phi\phi\right)_{\mathbf{3}}\left(\phi\phi\right)_{\mathbf{3}}\right)_{\mathbf{1}}
+f_5\left(\left(\phi\phi\right)_{\mathbf{3}'}\left(\phi\phi\right)_{\mathbf{3}'}\right)_{\mathbf{1}}\,,\\
\nonumber&&V(\rho)=M^2_{\rho}\left(\rho\rho\right)_{\mathbf{1}}+\mu_{\rho}\left(\rho\left(\rho\rho\right)_{\mathbf{2}}\right)_{\mathbf{1}}+f_6\left(\rho\rho\right)_{\mathbf{1}}\left(\rho\rho\right)_{\mathbf{1}}\\
\nonumber&&\qquad\quad+f_7\left(\rho\rho\right)_{\mathbf{1}'}\left(\rho\rho\right)_{\mathbf{1}'}+f_8\left(\left(\rho\rho\right)_{\mathbf{2}}\left(\rho\rho\right)_{\mathbf{2}}\right)_{\mathbf{1}}\,,\\
\nonumber&&V(\zeta,\phi)=f_9\zeta\left(\phi\left(\phi\phi\right)_{\mathbf{3}'}\right)_{\mathbf{1}}+f_{10}\zeta^2\left(\phi\phi\right)_{\mathbf{1}}\,,\\
\nonumber&&V(\zeta,\rho)=f_{11}\zeta^2\left(\rho\rho\right)_{\mathbf{1}}\,,\\
&&V(\phi,\rho)=\mu_{\phi}\left(\rho\left(\phi\phi\right)_{\mathbf{2}}\right)_{\mathbf{1}}+f_{12}\left(\rho\rho\right)_{\mathbf{1}}\left(\phi\phi\right)_{\mathbf{1}}+f_{13}\left(\left(\rho\rho\right)_{\mathbf{2}}\left(\phi\phi\right)_{\mathbf{2}}\right)_{\mathbf{1}}\,,
\end{eqnarray}
and
\begin{eqnarray}
\nonumber&&V(\chi)=M^2_{\chi}\left(\chi\chi\right)_{\mathbf{1}}+\mu_{\chi}\left(\chi\left(\chi\chi\right)_{\mathbf{3}}\right)_{\mathbf{1}}+g_1\left(\chi\chi\right)_{\mathbf{1}}\left(\chi\chi\right)_{\mathbf{1}}\\
\nonumber&&\qquad\quad+g_2\left(\left(\chi\chi\right)_{\mathbf{2}}\left(\chi\chi\right)_{\mathbf{2}}\right)_{\mathbf{1}}+g_{3}\left(\left(\chi\chi\right)_{\mathbf{3}}\left(\chi\chi\right)_{\mathbf{3}}\right)_{\mathbf{1}}+
g_{4}\left(\left(\chi\chi\right)_{\mathbf{3}'}\left(\chi\chi\right)_{\mathbf{3}'}\right)_{\mathbf{1}}\,,\\
\nonumber&&V(\varphi)=M^2_{\varphi}\left(\varphi\varphi\right)_{\mathbf{1}}+g_{5}\left(\varphi\varphi\right)_{\mathbf{1}}\left(\varphi\varphi\right)_{\mathbf{1}}+g_6\left(\left(\varphi\varphi\right)_{\mathbf{2}}\left(\varphi\varphi\right)_{\mathbf{2}}\right)_{\mathbf{1}}\\
\nonumber&&\qquad\quad+g_7\left(\left(\varphi\varphi\right)_{\mathbf{3}}\left(\varphi\varphi\right)_{\mathbf{3}}\right)_{\mathbf{1}}+g_8\left(\left(\varphi\varphi\right)_{\mathbf{3}'}\left(\varphi\varphi\right)_{\mathbf{3}'}\right)_{\mathbf{1}}\,,\\
\nonumber&&V(\chi,\varphi)=\mu_{\varphi}\left(\chi\left(\varphi\varphi\right)_{\mathbf{3}}\right)_{\mathbf{1}}+g_{9}\left(\chi\chi\right)_{\mathbf{1}}\left(\varphi\varphi\right)_{\mathbf{1}}+g_{10}\left(\left(\chi\chi\right)_{\mathbf{2}}\left(\varphi\varphi\right)_{\mathbf{2}}\right)_{\mathbf{1}}\\
&&\qquad\quad+g_{11}\left(\left(\chi\chi\right)_{\mathbf{3}}\left(\varphi\varphi\right)_{\mathbf{3}}\right)_{\mathbf{1}}+g_{12}\left(\left(\chi\chi\right)_{\mathbf{3}'}\left(\varphi\varphi\right)_{\mathbf{3}'}\right)_{\mathbf{1}}\,.
\end{eqnarray}
We start by analyzing the vacuum configuration:
\begin{equation}
\langle\chi\rangle=\left(0,0,v_{\chi}\right),\qquad \langle\varphi\rangle=\left(0,v_{\varphi},0\right)\,.
\end{equation}
The minimization conditions are:
\begin{eqnarray}
\nonumber&&\frac{\partial\mathcal{V}}{\partial\chi_3}=2v_{\chi}\left[M^2_{\chi}+2(g_1+4g_2)v^2_{\chi}+(g_9-2g_{10})v^2_{\varphi}\right]-\mu_{\varphi}v^2_{\varphi}=0\,,\\
&&\frac{\partial\mathcal{V}}{\partial\varphi_2}=2v_{\varphi}\left[M^2_{\varphi}+2(g_5+g_6+g_7)v^2_{\varphi}+(g_{9}-2g_{10})v^2_{\chi}-\mu_{\varphi}v_{\chi}\right]=0\,,
\end{eqnarray}
while $\left(\partial\mathcal{V}/\partial\chi_{1,2}\right)=0$ and $\left(\partial\mathcal{V}/\partial\varphi_{1,3}\right)=0$ are automatically satisfied. In a non-vanishing portion of the parameter space, these equations have non-trivial solution with non-vanishing $v_{\chi}$ and $v_{\varphi}$. The minimization equations for the vacuum configuration of $\langle\zeta\rangle=v_{\zeta}$, $\langle\phi\rangle=\left(0,\sqrt{2},1\right)v_{\phi}$ and $\langle\rho\rangle=\left(v_{\rho},0\right)$ read
\begin{eqnarray}
\nonumber&&\frac{\partial\mathcal{V}}{\partial\zeta}=2v_{\zeta}\left(M^2_{\zeta}+2f_1v^2_{\zeta}+3f_{10}v^2_{\phi}+f_{11}v^2_{\rho}\right)=0\,,\\
\nonumber&&\frac{\partial\mathcal{V}}{\partial\rho_1}=v_{\rho}\left[2M^2_{\rho}-3\mu_{\rho}v_{\rho}+4(f_6+f_8)v^2_{\rho}+2f_{11}v^2_{\zeta}+6f_{12}v^2_{\phi}\right]=0\,,\\
\nonumber&&\frac{\partial\mathcal{V}}{\partial\phi_2}=2\sqrt{2}\;v_{\phi}\left[M^2_{\phi}+\mu_{\phi}v_{\rho}+2(3f_2+4f_4)v^2_{\phi}+f_{10}v^2_{\zeta}+(f_{12}-f_{13})v^2_{\rho}\right]=0\,,\\
\label{eq:vev_neutrino}&&\frac{\partial\mathcal{V}}{\partial\phi_3}=2v_{\phi}\left[M^2_{\phi}-2\mu_{\phi}v_{\rho}+2(3f_2+4f_4)v^2_{\phi}+f_{10}v^2_{\zeta}+(f_{12}+2f_{13})v^2_{\rho}\right]=0\,.
\end{eqnarray}
The remaining minimum conditions $\left(\partial\mathcal{V}/\partial{\rho_2}\right)=0$ and $\left(\partial\mathcal{V}/\partial{\phi_1}\right)=0$ are always satisfied in this case. From the set of equations of Eq.~\eqref{eq:vev_neutrino}, we obtain the solution
\begin{eqnarray}
\nonumber&&v_{\rho}=\frac{\mu_{\phi}}{f_{13}},\qquad v^2_{\phi}=\frac{2f_1M^2_{\rho}-f_{11}M^2_{\zeta}-3f_1\mu_{\rho}v_{\rho}+\left[4f_1(f_6+f_8)-f^2_{11}\right]v^2_{\rho}}{3\left(f_{10}f_{11}-2f_{1}f_{12}\right)}\,,\\
&&v^2_{\zeta}=\frac{2f_{12}M^2_{\zeta}-2f_{10}M^2_{\rho}+3f_{10}\mu_{\rho}v_{\rho}+2\left[f_{11}f_{12}-2f_{10}(f_6+f_8)\right]v^2_{\rho}}{2\left(f_{10}f_{11}-2f_1f_{12}\right)}\,.
\end{eqnarray}
In addition, we only need to fine tune the parameters of the potential to satisfy
\begin{equation}
M^2_{\phi}+2\left(3f_2+4f_4\right)v^2_{\phi}+f_{10}v^2_{\zeta}+f_{12}v^2_{\rho}=0\,.
\end{equation}
Such restriction can be avoided by switching off the interaction potential $V(\phi,\rho)$ between $\phi$ and $\rho$, and this scenario could be naturally realized in supersymmetric dynamical completion, as shown in Refs.~\cite{Altarelli:2005yp,Altarelli:2005yx}. A rigorous explanation of this possibility is beyond the scope of the present work.


\newpage

\begin{thebibliography}{}


\bibitem{RS_model}
  L.~Randall and R.~Sundrum,
  Phys.\ Rev.\ Lett.\  {\bf 83}, 3370 (1999)
  [arXiv:hep-ph/9905221];
  L.~Randall and R.~Sundrum,
  Phys.\ Rev.\ Lett.\  {\bf 83}, 4690 (1999)
  [arXiv:hep-th/9906064].

\bibitem{bulkfields}
  H.~Davoudiasl, J.~L.~Hewett and T.~G.~Rizzo,
  Phys.\ Lett.\  B {\bf 473}, 43 (2000)
  [arXiv:hep-ph/9911262];
  A.~Pomarol,
  Phys.\ Lett.\  B {\bf 486}, 153 (2000)
  [arXiv:hep-ph/9911294];
  S.~Chang, J.~Hisano, H.~Nakano, N.~Okada and M.~Yamaguchi,
  Phys.\ Rev.\  D {\bf 62}, 084025 (2000)
  [arXiv:hep-ph/9912498];
  S.~J.~Huber and Q.~Shafi,
  Phys.\ Rev.\  D {\bf 63}, 045010 (2001)
  [arXiv:hep-ph/0005286].


\bibitem{hierarchy1_RS}
S.~J.~Huber and Q.~Shafi,
  Phys.\ Lett.\  B {\bf 498}, 256 (2001)
  [arXiv:hep-ph/0010195];
S.~J.~Huber,
  Nucl.\ Phys.\  B {\bf 666}, 269 (2003)
  [arXiv:hep-ph/0303183].


\bibitem{hierarchy2_RS}
K.~Agashe, G.~Perez and A.~Soni,
  Phys.\ Rev.\ Lett.\  {\bf 93}, 201804 (2004)
  [arXiv:hep-ph/0406101];
  Phys.\ Rev.\  D {\bf 71}, 016002 (2005)
  [arXiv:hep-ph/0408134].


\bibitem{Huber:2000fh}
  S.~J.~Huber and Q.~Shafi,
  Phys.\ Rev.\ D {\bf 63}, 045010 (2001)
  [arXiv:hep-ph/0005286];
  S.~J.~Huber, C.~A.~Lee and Q.~Shafi,
  Phys.\ Lett.\ B {\bf 531}, 112 (2002)
  [arXiv:hep-ph/0111465];
  C.~Csaki, J.~Erlich and J.~Terning,
  Phys.\ Rev.\ D {\bf 66}, 064021 (2002)
  [arXiv:hep-ph/0203034];
  J.~L.~Hewett, F.~J.~Petriello and T.~G.~Rizzo,
  JHEP {\bf 0209}, 030 (2002)
  [arXiv:hep-ph/0203091];
  M.~S.~Carena, E.~Ponton, J.~Santiago and C.~E.~M.~Wagner,
   Nucl.\ Phys.\ B {\bf 759}, 202 (2006)  [arXiv:hep-ph/0607106].  


\bibitem{Agashe:2003zs}
K.~Agashe, A.~Delgado, M.~J.~May and R.~Sundrum,
JHEP {\bf 0308}, 050 (2003)
[arXiv:hep-ph/0308036].


\bibitem{Agashe:2006at}
  K.~Agashe, R.~Contino, L.~Da Rold and A.~Pomarol,
Phys.\ Lett.\ B {\bf 641}, 62 (2006)  [arXiv:hep-ph/0605341].


\bibitem{Davoudiasl:2002ua}
  H.~Davoudiasl, J.~L.~Hewett and T.~G.~Rizzo,
  Phys.\ Rev.\ D {\bf 68}, 045002 (2003)
  [arXiv:hep-ph/0212279];
  M.~Carena, E.~Pont\'{o}n, T.~M.~P.~Tait and C.~E.~M.~Wagner,
  Phys.\ Rev.\ D {\bf 67}, 096006 (2003)
  [arXiv:hep-ph/0212307];
  M.~Carena, A.~Delgado, E.~Pont\'{o}n, T.~M.~P.~Tait and C.~E.~M.~Wagner,
  Phys.\ Rev.\ D {\bf 71}, 015010 (2005)
  [arXiv:hep-ph/0410344].


\bibitem{Cabrer:2010si}
  J.~A.~Cabrer, G.~von Gersdorff and M.~Quiros,
  Phys.\ Lett.\ B {\bf 697}, 208 (2011)
  [arXiv:1011.2205 [hep-ph]];
  J.~A.~Cabrer, G.~von Gersdorff and M.~Quiros,
  JHEP {\bf 1105}, 083 (2011)
  [arXiv:1103.1388 [hep-ph]].


\bibitem{Gherghetta:2000qt}
  T.~Gherghetta and A.~Pomarol,
  Nucl.\ Phys.\  B {\bf 586}, 141 (2000)
  [arXiv:hep-ph/0003129].


\bibitem{Agashe:2004ay}
  K.~Agashe, G.~Perez and A.~Soni,
  Phys.\ Rev.\ Lett.\  {\bf 93}, 201804 (2004)
  [arXiv:hep-ph/0406101];
  K.~Agashe, G.~Perez and A.~Soni,
  Phys.\ Rev.\   {\bf D71}, 016002 (2005)
  [arXiv:hep-ph/0408134];
  K.~Agashe, M.~Papucci, G.~Perez and D.~Pirjol,
  arXiv:hep-ph/0509117;
  Z.~Ligeti, M.~Papucci and G.~Perez,
  Phys.\ Rev.\ Lett.\  {\bf 97}, 101801 (2006)
  [arXiv:hep-ph/0604112].


\bibitem{Cacciapaglia:2007fw}
  G.~Cacciapaglia, C.~Csaki, J.~Galloway, G.~Marandella, J.~Terning and A.~Weiler,
  JHEP {\bf 0804}, 006 (2008)
  [arXiv:0709.1714 [hep-ph]].


\bibitem{Gedalia:2009ws}
  O.~Gedalia, G.~Isidori and G.~Perez,
  Phys.\ Lett.\ B {\bf 682}, 200 (2009)  [arXiv:0905.3264 [hep-ph]].  


\bibitem{Kitano:2000wr}
  R.~Kitano,
  Phys.\ Lett.\  {\bf B481}, 39 (2000)
  [arXiv:hep-ph/0002279].
  T.~P.~Cheng and L.~-F.~Li,
  Phys.\ Lett.\ B {\bf 502}, 152 (2001)  [arXiv:hep-ph/0101068].  


\bibitem{Moreau:2006np}
  G.~Moreau and J.~I.~Silva-Marcos,
  JHEP {\bf 0603}, 090 (2006)
  [arXiv:hep-ph/0602155];
  S.~Davidson, G.~Isidori and S.~Uhlig,
  Phys.\ Lett.\  B {\bf 663}, 73 (2008)
  [arXiv:0711.3376 [hep-ph]].

\bibitem{Agashe:2006iy}
  K.~Agashe, A.~E.~Blechman and F.~Petriello,
  Phys.\ Rev.\  {\bf D74}, 053011 (2006)
  [arXiv:hep-ph/0606021];
  K.~Agashe,
  Phys.\ Rev.\ D {\bf 80}, 115020 (2009)  [arXiv:0902.2400 [hep-ph]].  



\bibitem{Fitzpatrick:2007sa}
  A.~L.~Fitzpatrick, G.~Perez and L.~Randall,
Phys.\ Rev.\ Lett.\  {\bf 100}, 171604 (2008)  [arXiv:0710.1869 [hep-ph]]; 
  J.~Santiago,
JHEP {\bf 0812}, 046 (2008)  [arXiv:0806.1230 [hep-ph]].  


\bibitem{Chen:2008qg}
  M.-C.~Chen and H.~B.~Yu,
  Phys.\ Lett.\  {\bf B672}, 253 (2009)
  [arXiv:0804.2503 [hep-ph]];
  G.~Perez and L.~Randall,
  JHEP {\bf 0901}, 077 (2009)
  [arXiv:0805.4652 [hep-ph]].


\bibitem{Csaki:2009wc}
  C.~Csaki, G.~Perez, Z.~'e.~Surujon and A.~Weiler,
  Phys.\ Rev.\ D {\bf 81}, 075025 (2010)  [arXiv:0907.0474 [hep-ph]].  

\bibitem{Csaki:2008qq}
  C.~Csaki, C.~Delaunay, C.~Grojean and Y.~Grossman,
  JHEP {\bf 0810}, 055 (2008)
  [arXiv:0806.0356 [hep-ph]].


\bibitem{Altarelli:2005yp}
  G.~Altarelli and F.~Feruglio,
Nucl.\ Phys.\ B {\bf 720}, 64 (2005)  [arXiv:hep-ph/0504165].  


\bibitem{Altarelli:2008bg}
  G.~Altarelli, F.~Feruglio and C.~Hagedorn,
  JHEP {\bf 0803}, 052 (2008)  [arXiv:0802.0090 [hep-ph]].  
  T.~J.~Burrows and S.~F.~King,
  Nucl.\ Phys.\ B {\bf 835}, 174 (2010)  [arXiv:0909.1433 [hep-ph]].  



\bibitem{Chen:2009gy}
  M.~-C.~Chen, K.~T.~Mahanthappa and F.~Yu,
  Phys.\ Rev.\ D {\bf 81}, 036004 (2010)  [arXiv:0907.3963 [hep-ph]]; 
  A.~Kadosh and E.~Pallante,
  JHEP {\bf 1008}, 115 (2010)  [arXiv:1004.0321 [hep-ph]]; 
  A.~Kadosh and E.~Pallante,
   JHEP {\bf 1106}, 121 (2011)  [arXiv:1101.5420 [hep-ph]];  
  A.~Kadosh,
  arXiv:1303.2645 [hep-ph].  


\bibitem{delAguila:2010vg}
  F.~del Aguila, A.~Carmona and J.~Santiago,
JHEP {\bf 1008}, 127 (2010)  [arXiv:1001.5151 [hep-ph]];
  C.~Hagedorn and M.~Serone,
  JHEP {\bf 1202}, 077 (2012)  [arXiv:1110.4612 [hep-ph]];  
  JHEP {\bf 1110}, 083 (2011)  [arXiv:1106.4021 [hep-ph]].  


\bibitem{TBmix} P.~F.~Harrison, D.~H.~Perkins and W.~G.~Scott, Phys.\ Lett.\  B {\bf 530}, 167
(2002), [arXiv:hep-ph/0202074]; P.~F.~Harrison and W.~G.~Scott, Phys.\
Lett.\  B {\bf 535}, 163 (2002), [arXiv:hep-ph/0203209]; Z.~Z.~Xing, Phys.\
Lett.\  B {\bf 533}, 85 (2002), [arXiv:hep-ph/0204049]; X.~G.~He and A.~Zee,
Phys.\ Lett.\  B {\bf 560}, 87 (2003), [arXiv:hep-ph/0301092].


\bibitem{Abe:2011sj}
  K.~Abe {\it et al.}  [T2K Collaboration],
Phys.\ Rev.\ Lett.\  {\bf 107}, 041801 (2011)  [arXiv:1106.2822 [hep-ex]]; arXiv:1304.0841 [hep-ex].  


\bibitem{Adamson:2011qu}
  P.~Adamson {\it et al.}  [MINOS Collaboration],
Phys.\ Rev.\ Lett.\  {\bf 107}, 181802 (2011)  [arXiv:1108.0015 [hep-ex]].  


\bibitem{Abe:2011fz}
  Y.~Abe {\it et al.}  [DOUBLE-CHOOZ Collaboration],
  Phys.\ Rev.\ Lett.\  {\bf 108}, 131801 (2012)  [arXiv:1112.6353 [hep-ex]]; 
arXiv:1301.2948 [hep-ex].  


\bibitem{An:2012eh}
  F.~P.~An {\it et al.}  [DAYA-BAY Collaboration],
   Phys.\ Rev.\ Lett.\  {\bf 108}, 171803 (2012)  [arXiv:1203.1669 [hep-ex]];  
  Chin.\  Phys.\ C {\bf 37}, 011001 (2013)  [arXiv:1210.6327 [hep-ex]].


\bibitem{Ahn:2012nd}
  J.~K.~Ahn {\it et al.}  [RENO Collaboration],
  Phys.\ Rev.\ Lett.\  {\bf 108}, 191802 (2012)  [arXiv:1204.0626 [hep-ex]].  



\bibitem{Tortola:2012te}
 D.~V.~Forero, M.~Tortola and J.~W.~F.~Valle,
  Phys.\ Rev.\ D {\bf 86}, 073012 (2012)
  [arXiv:1205.4018 [hep-ph]].


\bibitem{Fogli:2012ua}
  G.~L.~Fogli, E.~Lisi, A.~Marrone, D.~Montanino, A.~Palazzo and A.~M.~Rotunno,
  Phys.\ Rev.\ D {\bf 86}, 013012 (2012)  [arXiv:1205.5254 [hep-ph]].

\bibitem{GonzalezGarcia:2012sz}
  M.~C.~Gonzalez-Garcia, M.~Maltoni, J.~Salvado and T.~Schwetz,
JHEP {\bf 1212}, 123 (2012)  [arXiv:1209.3023 [hep-ph]].  


\bibitem{Toorop:2011jn}
  R.~d.~A.~Toorop, F.~Feruglio and C.~Hagedorn,
Phys.\ Lett.\ B {\bf 703}, 447 (2011)  [arXiv:1107.3486 [hep-ph]];  
  R.~de Adelhart Toorop, F.~Feruglio and C.~Hagedorn,
  Nucl.\ Phys.\ B {\bf 858}, 437 (2012)  [arXiv:1112.1340 [hep-ph]].  


\bibitem{Ding:2012xx}
  G.~-J.~Ding,
  Nucl.\ Phys.\ B {\bf 862}, 1 (2012) [arXiv:1201.3279[hep-ph]]. 

\bibitem{King:2012in}
  S.~F.~King, C.~Luhn and A.~J.~Stuart,
  Nucl.\ Phys.\ B {\bf 867}, 203 (2013)  [arXiv:1207.5741 [hep-ph]].


\bibitem{Holthausen:2012wt}
  M.~Holthausen, K.~S.~Lim and M.~Lindner,
  Phys.\ Lett.\ B {\bf 721} 61 (2013)
  [arXiv:1212.2411 [hep-ph]]; 
  C.~S.~Lam,
 Phys.\ Rev.\ D {\bf 87}, 013001 (2013)  [arXiv:1208.5527 [hep-ph]].  


\bibitem{Hernandez:2012ra}
  S.~-F.~Ge, D.~A.~Dicus and W.~W.~Repko,
  Phys.\ Lett.\ B {\bf 702}, 220 (2011)
  [arXiv:1104.0602 [hep-ph]];
  S.~-F.~Ge, D.~A.~Dicus and W.~W.~Repko,
  Phys.\ Rev.\ Lett.\  {\bf 108}, 041801 (2012)
  [arXiv:1108.0964 [hep-ph]];
  D.~Hernandez and A.~Y.~Smirnov,
  Phys.\ Rev.\ D {\bf 86}, 053014 (2012)  [arXiv:1204.0445 [hep-ph]];  
  D.~Hernandez and A.~Y.~Smirnov,
  Phys.\ Rev.\ D {\bf 87}, 053005 (2013)
  [arXiv:1212.2149 [hep-ph]];  
  B.~Hu,
  Phys.\ Rev.\ D. {\bf 87}, 033002 (2013)
  [arXiv:1212.2819 [hep-ph]];  
  C.~S.~Lam,
  arXiv:1301.3121 [hep-ph].  


\bibitem{Lam:2013ng}
  C.~S.~Lam,
  arXiv:1301.1736 [hep-ph].  


\bibitem{Altarelli:2009gn}
  G.~Altarelli, F.~Feruglio and L.~Merlo,
  JHEP {\bf 0905}, 020 (2009)  [arXiv:0903.1940 [hep-ph]].  

\bibitem{Barger:1998ta}
  V.~D.~Barger, S.~Pakvasa, T.~J.~Weiler and K.~Whisnant,
  Phys.\ Lett.\ B {\bf 437}, 107 (1998)  [arXiv:hep-ph/9806387].  

\bibitem{Ding:2012wh}
  G.~-J.~Ding, S.~Morisi and J.~W.~F.~Valle,
  Phys.\ Rev.\ D {\bf 87,053013} (2013)  [arXiv:1211.6506 [hep-ph]].


\bibitem{Fritzsch:1995dj}
  H.~Fritzsch and Z.~-Z.~Xing,
  Phys.\ Lett.\ B {\bf 372}, 265 (1996)  [arXiv:hep-ph/9509389]; Phys.\ Lett.\ B {\bf 440}, 313 (1998)  [arXiv:hep-ph/9808272];  Phys.\ Rev.\ D {\bf 61}, 073016 (2000)  [arXiv:hep-ph/9909304].

\bibitem{Xing:2011at}
  Z.~-Z.~Xing,
  Chin.\ Phys.\ C {\bf 36}, 101 (2012)
  [arXiv:1106.3244 [hep-ph]].


\bibitem{Xing:2012ej}
  Z.~-Z.~Xing,
  Chin.\ Phys.\ C {\bf 36}, 281 (2012)  [arXiv:1203.1672 [hep-ph]].  


\bibitem{SeeSaw}
P. Minkowski,
  {\it Phys. Lett.} B {\bf 67} 421  (1977);
T.  Yanagida,
  in {\it Proc. of Workshop on Unified Theory and Baryon number in the Universe}, eds. O. Sawada and A. Sugamoto, KEK, Tsukuba, (1979) p.95;
M. Gell-Mann, P. Ramond and R. Slansky,
  in {\it Supergravity}, eds P.  van Niewenhuizen and D. Z. Freedman (North Holland, Amsterdam 1980) p.315;
P.  Ramond,
  {\it Sanibel talk}, retroprinted as hep-ph/9809459;
S. L. Glashow,
  in{\it Quarks and Leptons}, Carg\`ese lectures, eds M. L\'evy, (Plenum, 1980, New York) p. 707;
R. N. Mohapatra and G. Senjanovi\'c,
  {\it Phys. Rev.  Lett.} {\bf 44}, 912 (1980);
J.~Schechter and J.~W.~F.~Valle,
  Phys.\ Rev.\  D {\bf 22} (1980) 2227;
  Phys.\ Rev.\  D {\bf 25} (1982) 774.


\bibitem{Huber:2002gp}
  S.~J.~Huber and Q.~Shafi,
  Phys.\ Lett.\ B {\bf 544}, 295 (2002)  [arXiv:hep-ph/0205327]. 


\bibitem{Huber:2003sf}
  S.~J.~Huber and Q.~Shafi,
  Phys.\ Lett.\ B {\bf 583}, 293 (2004)  [arXiv:hep-ph/0309252].  



\bibitem{Grossman:1999ra}
  Y.~Grossman and M.~Neubert,
  Phys.\ Lett.\ B {\bf 474}, 361 (2000)  [arXiv:hep-ph/9912408]; 
  S.~J.~Huber and Q.~Shafi,
  Phys.\ Lett.\ B {\bf 512}, 365 (2001)  [arXiv:hep-ph/0104293];  
  G.~Moreau and J.~I.~Silva-Marcos,
  JHEP {\bf 0601}, 048 (2006)  [arXiv:hep-ph/0507145].  


\bibitem{Agashe:2008fe}
  K.~Agashe, T.~Okui and R.~Sundrum,
  Phys.\ Rev.\ Lett.\  {\bf 102}, 101801 (2009)  [arXiv:0810.1277 [hep-ph]].  


\bibitem{Memenga:2013vc}
  N.~Memenga, W.~Rodejohann and H.~Zhang,
  Phys.\ Rev.\ D {\bf 87}, 053021 (2013)
  [arXiv:1301.2963 [hep-ph]].


\bibitem{Lam:2008sh}
  C.~S.~Lam,
  Phys.\ Rev.\ D {\bf 78}, 073015 (2008)  [arXiv:0809.1185 [hep-ph]].


\bibitem{Ding:2011cm}
  G.~-J.~Ding, L.~L.~Everett and A.~J.~Stuart,
  Nucl.\ Phys.\ B {\bf 857}, 219 (2012)  [arXiv:1110.1688 [hep-ph]].  


\bibitem{Fritzsch:1989qm}
  H.~Fritzsch and J.~Plankl,
  Phys.\ Lett.\ B {\bf 237}, 451 (1990).



\bibitem{Blanke:2008zb}
  M.~Blanke, A.~J.~Buras, B.~Duling, S.~Gori and A.~Weiler,
  JHEP {\bf 0903}, 001 (2009)  [arXiv:0809.1073 [hep-ph]];  
  M.~Blanke, A.~J.~Buras, B.~Duling, K.~Gemmler and S.~Gori,
  JHEP {\bf 0903}, 108 (2009)  [arXiv:0812.3803 [hep-ph]];  
  S.~Casagrande, F.~Goertz, U.~Haisch, M.~Neubert, T.~Pfoh and ,
  JHEP {\bf 0810}, 094 (2008)
  [arXiv:0807.4937 [hep-ph]].


\bibitem{Breitenlohner:1982jf}
  P.~Breitenlohner and D.~Z.~Freedman,
  Annals Phys.\  {\bf 144}, 249 (1982).

\bibitem{Luty:2004ye}
  M.~A.~Luty and T.~Okui,
  JHEP {\bf 0609}, 070 (2006)
  [arXiv:hep-ph/0409274].


\bibitem{Cacciapaglia:2006mz}
  G.~Cacciapaglia, C.~Csaki, G.~Marandella and J.~Terning,
  JHEP {\bf 0702}, 036 (2007)  [arXiv:hep-ph/0611358];  


\bibitem{Davoudiasl:2005uu}
  H.~Davoudiasl, B.~Lillie and T.~G.~Rizzo,
  JHEP {\bf 0608}, 042 (2006)  [arXiv:hep-ph/0508279]; 
  G.~Cacciapaglia, C.~Csaki, G.~Marandella and J.~Terning,
  JHEP {\bf 0702}, 036 (2007)  [arXiv:hep-ph/0611358];  
  L.~Vecchi,
  JHEP {\bf 1111}, 102 (2011)  [arXiv:1012.3742 [hep-ph]];  
  A.~D.~Medina and E.~Ponton,
  JHEP {\bf 1106}, 009 (2011)  [arXiv:1012.5298 [hep-ph]];  
  J.~A.~Cabrer, G.~von Gersdorff and M.~Quiros,
  Phys.\ Rev.\ D {\bf 84}, 035024 (2011)  [arXiv:1104.3149 [hep-ph]]; 
  P.~R.~Archer,
  JHEP {\bf 1209}, 095 (2012)  [arXiv:1204.4730 [hep-ph]]; 
  M.~Frank, N.~Pourtolami and M.~Toharia,
  arXiv:1301.7692 [hep-ph];  
  R.~Malm, M.~Neubert, K.~Novotny and C.~Schmell,
  arXiv:1303.5702 [hep-ph].  


\bibitem{Carena:2006bn}
  M.~S.~Carena, E.~Ponton, J.~Santiago and C.~E.~M.~Wagner,
 Nucl.\ Phys.\ B {\bf 759}, 202 (2006)  [arXiv:hep-ph/0607106];  
  M.~E.~Albrecht, M.~Blanke, A.~J.~Buras, B.~Duling and K.~Gemmler,
JHEP {\bf 0909}, 064 (2009)  [arXiv:0903.2415 [hep-ph]].  


\bibitem{Buras:2009ka}
  A.~J.~Buras, B.~Duling and S.~Gori,
  JHEP {\bf 0909}, 076 (2009)  [arXiv:0905.2318 [hep-ph]];  


\bibitem{Barry:2010yk}
  J.~Barry and W.~Rodejohann,
  Nucl.\ Phys.\ B {\bf 842}, 33 (2011)
  [arXiv:1007.5217 [hep-ph]].


\bibitem{Fogli:2008cx}
  G.~L.~Fogli, E.~Lisi, A.~Marrone, A.~Palazzo and A.~M.~Rotunno,
  arXiv:0809.2936 [hep-ph].


\bibitem{WMAP2}
  E.~Komatsu {\it et al.}  [WMAP Collaboration],
  Astrophys.\ J.\ Suppl.\  {\bf 180}, 330 (2009)
  [arXiv:0803.0547 [astro-ph]].


\bibitem{acbar07}
  C.~L.~Reichardt, 
  {\it et al.}, [ACBAR Collaboration],
  Astrophys.\ J.\  {\bf 694}, 1200 (2009)  [arXiv:0801.1491 [astro-ph]].


\bibitem{vsa}
  C.~Dickinson, 
  {\it et al.}, [VSA Collaboration],
  Mon.\ Not.\ Roy.\ Astron.\ Soc.\  {\bf 353} (2004) 732  [astro-ph/0402498].


\bibitem{cbi}
  A.~C.~S.~Readhead, 
  {\it et al.}, [CBI Collaboration],
  Astrophys.\ J.\  {\bf 609}, 498 (2004)  [astro-ph/0402359].


\bibitem{boom03}
  C.~J.~MacTavish, 
  {\it et al.}, [BOOMERANG Collaboration],
  Astrophys.\ J.\  {\bf 647}, 799 (2006)  [astro-ph/0507503].


\bibitem{Tegmark}
  M.~Tegmark {\it et al.}  [SDSS Collaboration],
  Phys.\ Rev.\ D {\bf 74}, 123507 (2006)  [astro-ph/0608632].


\bibitem{astier}
  P.~Astier {\it et al.}  [SNLS Collaboration],
  Astrophys.\  {\bf 447}, 31 (2006)  [astro-ph/0510447].


\bibitem{bao}
  D.~J.~Eisenstein {\it et al.}  [SDSS Collaboration],
  Astrophys.\ J.\  {\bf 633}, 560 (2005)  [astro-ph/0501171].



\bibitem{Ly1}
  P.~McDonald {\it et al.}  [SDSS Collaboration],
  Astrophys.\ J.\ Suppl.\  {\bf 163}, 80 (2006)  [astro-ph/0405013];
  P.~McDonald {\it et al.}  [SDSS Collaboration],
  Astrophys.\ J.\  {\bf 635}, 761 (2005)  [astro-ph/0407377].



\bibitem{Ade:2013zuv}
  P.~A.~R.~Ade {\it et al.}  [Planck Collaboration],
  arXiv:1303.5076 [astro-ph.CO].  


\bibitem{Xing:2007fb}
  Z.~-z.~Xing, H.~Zhang and S.~Zhou,
  Phys.\ Rev.\ D {\bf 77}, 113016 (2008)
  [arXiv:0712.1419 [hep-ph]]; Phys.\ Rev.\ D {\bf 86}, 013013 (2012)
  [arXiv:1112.3112 [hep-ph]].



\bibitem{Feldman:2012qt}
  G.~J.~Feldman, J.~Hartnell and T.~Kobayashi,
  arXiv:1210.1778 [hep-ex].  


\bibitem{Beringer:1900zz}
  J.~Beringer {\it et al.}  [Particle Data Group Collaboration],
  Phys.\ Rev.\ D {\bf 86}, 010001 (2012).



\bibitem{Lam:2008rs}
  C.~S.~Lam,
  Phys.\ Rev.\ Lett.\  {\bf 101}, 121602 (2008)
  [arXiv:0804.2622 [hep-ph]];
  C.~S.~Lam,
  Phys.\ Rev.\ D {\bf 78}, 073015 (2008)
  [arXiv:0809.1185 [hep-ph]];
  C.~S.~Lam,
  Phys.\ Rev.\ D {\bf 83}, 113002 (2011)
  [arXiv:1104.0055 [hep-ph]].



\bibitem{Lam:2007qc}
  C.~S.~Lam,
  Phys.\ Lett.\ B {\bf 656}, 193 (2007)
  [arXiv:0708.3665 [hep-ph]].


\bibitem{Ding:2009iy}
  G.~-J.~Ding,
  Nucl.\ Phys.\ B {\bf 827}, 82 (2010)  [arXiv:0909.2210 [hep-ph]];  
  Nucl.\ Phys.\ B {\bf 846}, 394 (2011)  [arXiv:1006.4800 [hep-ph]].  


\bibitem{progress}
Work in progress.


\bibitem{Ma:2005pd}
  E.~Ma,
   Phys.\ Lett.\ B {\bf 632}, 352 (2006)  [arXiv:hep-ph/0508231].  

\bibitem{Hagedorn:2006ug}
  C.~Hagedorn, M.~Lindner and R.~N.~Mohapatra,
JHEP {\bf 0606}, 042 (2006)  [arXiv:hep-ph/0602244].  

\bibitem{Bazzocchi:2009pv}
  F.~Bazzocchi, L.~Merlo and S.~Morisi,
  Nucl.\ Phys.\ B {\bf 816}, 204 (2009)  [arXiv:0901.2086 [hep-ph]];
Phys.\ Rev.\ D {\bf 80}, 053003 (2009)  [arXiv:0902.2849 [hep-ph]].  


\bibitem{Hagedorn:2010th}
  C.~Hagedorn, S.~F.~King and C.~Luhn,
  JHEP {\bf 1006}, 048 (2010)  [arXiv:1003.4249 [hep-ph]];
  S.~F.~King and C.~Luhn,
  JHEP {\bf 1109}, 042 (2011)  [arXiv:1107.5332 [hep-ph]].



\bibitem{Altarelli:2005yx}
  G.~Altarelli and F.~Feruglio,
  Nucl.\ Phys.\ B {\bf 741}, 215 (2006)
  [hep-ph/0512103].


\end{thebibliography}
\end{document}